\documentclass[journal]{IEEEtran}
\usepackage[utf8]{inputenc}
\usepackage{algorithm2e}
\usepackage{mathrsfs}  
\usepackage{pbox}

\usepackage{cite}
\usepackage{amssymb}
\usepackage{diagbox}
\usepackage{url}
\usepackage{footnote}
\usepackage{hhline}
\usepackage{dsfont}
\ifCLASSINFOpdf
  \usepackage[pdftex]{graphicx}
\else
\fi
\usepackage[]{float}
\usepackage[caption = false]{subfig}

\newcommand\numberthis{\addtocounter{equation}{1}\tag{\theequation}}

\usepackage[cmex10]{amsmath}
\usepackage{amsthm}

\usepackage{amsfonts}
\usepackage{color}
\usepackage[space]{grffile}
\usepackage{todonotes}
\usepackage{microtype}
\usepackage{soul}
\setstcolor{red}
\usepackage{array,multirow}
\newcommand\cc{\textcolor{black}}
\newcommand\ccc{\textcolor{black}}

\begin{document}
\title{Hyperspectral Image Unmixing with Endmember Bundles and Group Sparsity Inducing Mixed Norms}

\author{Lucas~Drumetz,~\IEEEmembership{Member,~IEEE},  
     Travis~R.~Meyer,~  
         Jocelyn~Chanussot,~\IEEEmembership{Fellow,~IEEE} , 
        Andrea~L.~Bertozzi,~\IEEEmembership{Member,~IEEE},
         and~Christian~Jutten,~\IEEEmembership{Fellow,~IEEE}
       % <-this % stops a space
\thanks{L. Drumetz is with IMT Atlantique, Lab-STICC, UBL, Technopôle Brest-Iroise CS 83818, 29238 Brest Cedex 3, France (e-mail: lucas.drumetz@imt-atlantique.fr).}
\thanks{T. R. Meyer is with SmartKYC Ltd., Los Angeles, CA, USA (e-mail: travis.meyer@smartkyc.com).}
\thanks{J. Chanussot and C. Jutten are Univ. Grenoble Alpes, CNRS, Grenoble INP*, GIPSA-lab, 38000 Grenoble, France.  * Institute of Engineering Univ. Grenoble Alpes (e-mail: \{jocelyn.chanussot,christian.jutten\}@gipsa-lab.grenoble-inp.fr).}
\thanks{A. L. Bertozzi is with the Department of Mathematics at the University of California, Los Angeles, 90095 Los Angeles, CA, USA (e-mail: \{tmeyer;bertozzi\}@math.ucla.edu).}
%\thanks{G. Tochon and J. Chanussot are with the Grenoble Institute of Technology, GIPSA-lab, Image and Signal Department (DIS), F-38402 Saint Martin d'Heres Cedex (e-mail: \{guillaume.tochon;jocelyn.chanussot\}@gipsa-lab.fr, @gipsa-lab.fr)}

\thanks{This work was supported by the European Research Council FP7/2007–2013, under Grant no.320684 (CHESS project), as
well as projects ANR-16-ASTR-0027-01 APHYPIS, by CNRS PICS 263484, and by NSF grant DMS-1118971, NSF grant DMS-1417674, ONR
grant N000141210838, UC Lab Fees Research grant 12-LR-
236660. L. Drumetz was also supported by a Campus France outgoing postdoctoral mobility grant, PRESTIGE-2016-4 0006.}

\thanks{This paper has supplementary downloadable material available at http://ieeexplore.ieee.org., provided by the authors, which includes complementary results on the AVIRIS Cuprite dataset.}}

\ifCLASSOPTIONcaptionsoff
  \newpage
\fi

\maketitle

\begin{abstract}
Hyperspectral images provide much more information than conventional imaging techniques, allowing a precise identification of the materials in the observed scene, but because of the limited spatial resolution, the observations are usually mixtures of the contributions of several materials. The spectral unmixing problem aims at recovering the spectra of the pure materials of the scene (endmembers), along with their proportions (abundances) in each pixel. In order to deal with the intra-class variability of the materials and the induced spectral variability of the endmembers, several spectra per material, constituting endmember bundles, can be considered. However, the usual abundance estimation techniques do not take advantage of the particular structure of these bundles, organized into groups of spectra. In this paper, we propose to use group sparsity by introducing mixed norms in the abundance estimation optimization problem. In particular, we propose a new penalty which simultaneously enforces group and within group sparsity, to the cost of being nonconvex. All the proposed penalties are compatible with the abundance sum-to-one constraint, which is not the case with traditional sparse regression. We show on simulated and real datasets that well chosen penalties can significantly improve the unmixing performance compared \cc{to classical sparse regression techniques or} to the naive bundle approach.
\end{abstract}


% \fi
\IEEEpeerreviewmaketitle
\section{Introduction}
\IEEEPARstart{H}{yperspectral} imaging, also known as imaging spectroscopy, is a technique which allows to acquire  information in each pixel under the form of a spectrum of reflectance or radiance values for many --typically hundreds of-- narrow and contiguous wavelengths of the electromagnetic spectrum, usually (but not exclusively) in the visible and infra-red domains~\cite{Bioucas2013}. The fine spectral resolution of these images allow an accurate identification of the materials present in the scene, since two materials usually have distinct spectral profiles. However, this identification is made harder by the relatively low spatial resolution (significantly lower than panchromatic, color or even multispectral images). Therefore, many pixels are acquired with several materials in the field of view of the sensor, and the resulting observed signature is a mixture of the contributions of these materials. Spectral Unmixing (SU) is then a source separation problem whose goal is to recover the signatures of the pure materials of the scene (called \emph{endmembers}), and to estimate their relative proportions (called \emph{fractional abundances}) in each pixel of the image~\cite{Keshava2002}.

Usually, since the abundances are meant to be interpreted as proportions, they are required to be positive and to sum to one in each pixel. Another classical assumption made for SU is that the mixture of the contributions of the materials is linear, leading to the standard Linear Mixing Model (LMM)~\cite{bioucas2012,6678258}. Let us denote a hyperspectral image by $\mathbf{X}\in\mathbb{R}^{L\times N}$, gathering the pixels $\mathbf{x}_{k}\in\mathbb{R}^{L}$ ($k = 1,...,N$) in its columns, where $L$ is the number of spectral bands, and $N$ is the number of pixels in the image. The signatures $\mathbf{s}_{p},\ p = 1,...,P$ of the $P$ endmembers considered for the unmixing are gathered in the columns of a matrix $\mathbf{S}\in \mathbb{R}^{L\times P}$. The abundance coefficients $a_{pk}$ for each pixel $k = 1,...,N$ and material $p = 1,...,P$ are stored in the matrix $\mathbf{A}\in\mathbb{R}^{P\times N}$. With these notations, the LMM writes:
\begin{equation}
\mathbf{x}_{k} = \sum_{p = 1}^{P} a_{pk} \mathbf{s}_{p} + \mathbf{e}_{k},
\label{LMM}
\end{equation}
where $\mathbf{e}_{k}$ is an additive noise (usually assumed to be Gaussian distributed). Eq.~(\ref{LMM}) can be rewritten in a matrix form for the whole image:
\begin{equation}
\mathbf{X} = \mathbf{S}\mathbf{A} + \mathbf{E},
\label{LMM_mat}
\end{equation}
where $\mathbf{E}\in \mathbb{R}^{L\times N}$ comprises all the noise values. We keep in mind the constraints on the abundances: Abundance Nonnegativity Constraint (ANC) $a_{pk} \geq 0\ \forall (p,k)$ and the Abundance Sum-to-one Constraint (ASC) $\sum_{p=1}^{P} a_{pk} = 1, \forall k$. This model is considered physically realistic when each ray of light reaching the sensor has interacted with no more than one material on the ground (the so-called checkerboard configuration)~\cite{bioucas2012}. A nice property of the LMM is that, combined with the constraints on the abundances, it provides a strong geometrical structure to the problem: indeed the ANC and ASC force the abundances to lie in the unit simplex with $P$ vertices, which we denote by $\Delta_{P}$. Since the data are obtained via a linear transformation of the abundances, they are also constrained to lie in a simplex, whose vertices are the endmembers. The actual signal  subspace is then a $P-1$-dimensional subspace, embedded in the ambient space.

The typical blind SU chain comprises three steps:
\begin{itemize}
\item Estimating the number of endmembers to consider. This is a very hard and ill-posed problem in itself (because there is no such thing as an optimal number of endmembers in real data, among other reasons) and many algorithms have been considered in the community to try to obtain a good estimate~\cite{Robin2015}.
\item Extracting the spectra corresponding to the endmembers, a procedure referred to as endmember extraction. Then again, many Endmember Extraction Algorithms (EEAs) exist in the litterature to tackle this problem, with various assumptions, the main one being the presence in the data of pure pixels, i.e. pixels in which only one material of interest is present~\cite{Plaza2004}. These algorithms try to exploit the geometry of the problem by looking for extreme pixels in the data, which are the endmembers if the LMM holds. A popular EEA using the pure pixel assumption is the Vertex Component Analysis (VCA)~\cite{Nascimento2005}.
\item Finally, estimating the abundances using the data and the extracted endmembers. This step is usually carried out by solving a constrained optimization problem:
\begin{equation}
\underset{\mathbf{A}\geq \mathbf{0}, \ \mathds{1}_{P}^{\top}\mathbf{A} = \mathds{1}_{N}^{\top}}{\textrm{arg min}} \ ||\mathbf{X}-\mathbf{SA}||_{F}^{2},
\label{FCLSU}
\end{equation}
where $\mathds{1}_{\cdot}$ denotes a vector of ones whose size is given in index, and $||\cdot||_{F}$ is the Frobenius norm. Solving this problem is often referred to as Fully Constrained Least Squares Unmixing (FCLSU)~\cite{Heinz2001}.
\end{itemize}
This unmixing chain is now standard and has been used with success over the last two decades. However, the main two limitations of this approach have been identified as nonlinear effects, and endmember variability. Nonlinear interactions may occur if light reaching the sensor has interacted with more than one material on the ground (e.g. in tree canopies, or in particulate media). These physical considerations have spanned a number of approaches taking some nonlinear effects into consideration; see~\cite{heylen2014,6678284} for reviews on that subject. The other limitation, termed spectral variability, is simply based on the consideration that all materials have a certain intra-class variability, and therefore cannot be represented by a single signature, as in the conventional LMM~\cite{zare2014, drumetz2016endmember}. Note that even though both phenomena are usually considered separately, some approaches attempt to address both simultaneously~\cite{7508937}. Recently, a formal link between models addressing these two limitations was derived in~\cite{8022979}.

Causes of variability of the materials from one pixel to the other include changing illumination conditions locally in the scene, especially because of a nonflat topography, which will change the local incident angle of the light and the viewing angle of the sensor. Also, physico-chemical changes in the composition of the materials induce modifications on their spectra. For example, the concentration of chlorophyll in grass or soil moisture content can change the signatures of these two materials. Several approaches have been designed for variability aware unmixing, that is to correct the abundances due to the intra-class variations of the endmembers. The existing methods could be basically divided into two groups: dictionary or bundle-based approaches~\cc{\cite{bateson2000endmember, somers2012, xu2015image,xu2018regional}}, which try to model an endmember by a certain number of instances of each material, and model-based approaches, which define a specific model for the variation of the endmembers, be it computational~\cite{PLMM,8264812,7217817} or more physics-inspired~\cite{drumetztip}. Even though most methods consider spectral variability in the spatial domain, within a single image, studies dealing with temporal endmember variability have also been recently conducted~\cite{henrot2016,thouvenin2016online}.

This paper focuses on variability in the spatial domain using bundle-based approaches. Even though they lack the interpretability of their model based counterparts, they can still fit the framework of the LMM, while correcting the abundances with less assumptions on the data. The idea behind spectral bundles is to find a way to extract a dictionary of different instances of the materials directly from the data. Then one can simply replace the matrix $\mathbf{S}$ in Eq.~(\ref{LMM_mat}) by a dictionary made of spectral bundles $\mathbf{B}\in \mathbb{R}^{L\times Q}$, where $Q>P$ is the total number of endmember candidates. Since this dictionary (and the abundance coefficients as well) is organized into $P$ groups, and since $Q$ can be relatively large, it makes sense to consider that only a few atoms of the bundle dictionary are going to be used in  each pixel. Moreover, a now common assumption in SU is that a few materials are active in each pixel, out of the $P$ considered materials. These two assumptions, though similar, are not equivalent. In any case, they justify the use of sparsity in the SU problem. For instance, sparse regression using libraries of endmembers (i.e. semi-supervised unmixing) has been investigated in~\cite{Iordache2011} and exploited by many authors thereafter. It is interesting to note that sparse unmixing, when performed using convex optimization, and more precisely $\mathcal{L}_{1}$ norm regularization, is not compatible with the ASC. We will come back to this issue in the next section. When dealing with coefficients organized in groups, a small number of which are supposed to be active, one may resort to approaches similar to the so-called group LASSO (for Least Absolute Shrinkage and Selection Operator)~\cite{meier2008group}. In SU, both approaches have been combined in semi-supervised unmixing in~\cite{iordache2011hyperspectral}, when the dictionary can be organized into clusters of signatures.

The contributions of this paper are threefold. We first show that when using the classical FCLSU to estimate the abundances of the materials in the context of endmember bundles, it is possible to recover pixel-wise endmembers representing variability effects more subtly than with the signatures of the bundle dictionary, and we provide a simple geometrical interpretation for them. Second, further elaborating on the considerations and results of the conference paper \cite{7532746},  we improve the abundance estimation when bundles are used by considering three ``social" sparsity inducing penalties~\cite{kowalski2009sparsity, kowalski2013social}, whose interest we show by comparing their performance on simulated and real data to the classical \cc{sparse regression techniques (not using the group information)} and the naive bundle approach. Third, while two of those penalties have already been used in various signal processing problems, the last one (a mixed fractional $\mathcal{L}_{1,\frac{a}{b}}$ quasinorm with $a,b$ integers) is new to the best of our knowledge, and presents remarkable properties, while raising interesting technical challenges in its optimization.

The remainder of this paper is organized as follows: Section~\ref{bundles} presents in more detail the basic procedure to obtain a bundle dictionary from the image data, based on~\cite{somers2012}. While this is not the core of our method, it still serves as a basis for blind SU with bundles, and we present it here for the reader's convenience. We also propose a new geometric interpretation to SU using spectral bundles with \cc{any algorithm which estimates one abundance coefficient for every atom of the dictionary (e.g. all the algorithms considered in this study)}. Section~\ref{proposed} introduces the proposed sparsity inducing penalties, and details how to handle the resulting optimization problem for abundance estimation in each case. Section~\ref{results} tests the different penalties on synthetic and real data, and compares them to \cc{classical sparse regression techniques and to} the standard bundle unmixing approach. Finally, Section~\ref{conclusion} gathers some concluding remarks and possible future research avenues.

\section{Dealing with spectral variability with endmember bundles}
\label{bundles}

\subsection{Automated Endmember Bundles}

Extracting endmember bundles from the image data is a procedure whose goal is to obtain various instances of each endmember, still in a blind unmixing setting. We present and will use the so-called Automated Endmember Bundles (AEB) approach of~\cite{somers2012}. It is conceptually very simple and based on the following idea: since running a classical EEA once on the data provides the extreme points of the dataset, several instances of each material could be obtained by running an EEA on several subsets of the data. Then, one can randomly sample $m$ subsets of the data of size $N$ (pixels), and run an EEA on each subset in order to obtain $P$ endmember candidates, and therefore $Q = mP$ signatures in total. It is possible that for a given run, an EEA extracts several signatures corresponding to the same material. Thus, it is not guaranteed that each endmember is represented by exactly $m$ instances. Some rare endmembers could also appear. Usually, however, the number of desired bundles is set as the estimated number of endmembers for the whole image. In addition, even in a perfect case, the instances for each endmember are not aligned, i.e. the ordering of the extracted endmembers in different subsets is not the same, due to the stochasticity of the EEAs. Thus, a clustering step is required to group the signatures into bundles. In~\cite{somers2012}, the clustering is performed using the k-means algorithm, with the spectral angle as a similarity measure. The spectral angle has the nice property of being insensitive to scalings of the input vectors, which is a well-known effect of spectral variability related to changing illumination conditions~\cite{drumetztip}. \cc{More recent bundle extraction methods~\cite{xu2015image,xu2018regional} also incorporate spatial information to select endmember candidates in spatially homogeneous areas of the image.}

This idea of extracting several endmember candidates in different parts of the image is related to the recent Local Spectral Unmixing approaches, which handle variability by assuming it is less important in \emph{local} (as opposed to \emph{random} here) subsets of the image. In this type of methods, the whole unmixing process is entirely carried out in each (connected) subset. Some examples of local unmixing include~\cite{Goenaga2013, canham2011, Veganzones2014, tochon2016}. The main difference here is that we extract endmembers in random subsets, before gathering all those signatures into a common pool organized into groups, from which we estimate abundances for the whole image, not in local subsets.

%% Explain the used clustering method instead of k means. For now it is k-means on the principal components. However, more robust graph based techniques should be considers: real spectral clustering, Nystrom extension or graph MBO.
In any case, the result of the clustering is a dictionary $\mathbf{B}$ of endmember candidates. The clustering step defines a group structure\footnote{This denomination should not be understood in the sense of the algebraic structure. Instead, this simply means that the endmembers (and hence the abundance coefficients) are grouped into $P$ clusters, which we call groups here.} on the abundance coefficients and on the dictionary. We denote this group structure by $G$, and each group $G_{i}$, $i = 1,...,P$ contains $m_{G_{i}}$ signatures, so that representative $j$ of group $G_{i}$  in the dictionary is denoted as $\mathbf{b}_{G_{i},{j}}$. Then there are $Q = \sum_{i=1}^{P}m_{G_{i}}$ columns in $\mathbf{B}\in\mathbb{R}^{L\times Q}$.

Let us note that obviously, the quality of the unmixing is going to be very dependent on the quality of the bundle itself. In some cases, a spectral library of endmembers incorporating some endmember variability may be available directly. In such a case, the methods we describe next still apply, and the unmixing paradigm becomes semi-supervised, as in the sparse unmixing approaches~\cite{Iordache2011}.
\subsection{Abundance Estimation}

Now that the bundles are available, one has to estimate the abundances of the different materials, which can be done in multiple ways. The Fisher Discriminant Nullspace (FDN) approach~\cite{Jin2010} is a dimension reduction technique based on finding a linear transformation which maps the data to a subspace where the intra class variability of the bundles is minimized, while their inter class variability is maximized. The unmixing is then carried out in this new subspace, e.g. using the centroids of each class a the endmembers.

Multiple Endmember Spectral Mixture Analysis (MESMA)~\cite{Roberts1998} is a technique which aims at selecting in each pixel of the HSI the best endmember candidate for each material in terms of data fit. To do that,~Problem (\ref{FCLSU}) is solved using all possible combinations of (a small number of) candidate endmembers as the endmember matrix. However, the problem thus becomes combinatorial and is computationally expensive. \cc{Besides, MESMA implicitly assumes that a single instance of each endmember is active in each pixel, which may limit the flexibility of the model.}

Machine learning-inspired approaches have also been designed, using the fact that ``training data" can be constructed by simulating mixtures of the different instances of the materials  in various proportions \cite{mianji2011,uezato2016}.

Last, but not least, one can simply replace the endmember matrix $\mathbf{S}$ with the bundle matrix $\mathbf{B}$ in Eq.~(\ref{FCLSU}) and solve for the new abundance matrix in the exact same way, via constrained least squares. This approach is the most straightworward, but differs from the ones mentioned above insofar as several endmembers from each bundle can be simultaneously active, and summing the individual contributions of each material in each pixel is required to recover the abundances of the materials. This can be seen as a drawback, which makes the interpretation of the results harder. Of course, when the dictionary is not extracted from the data, and each instance represents a particular physical configuration of a material, this could indeed be an issue if the goal is to associate each pixel with a given physical parameter, though it is technically possible that several configurations of the same material can be present at the same time in a given pixel. \cc{Allowing several instances of a material to be active in a single pixel correspond to situations where several physico-chemical variants of the same material coexist in the field of view of the sensor for a single pixel (e.g. burnt and healthy grass)}. Here, we show that this solution is actually theoretically more powerful than considering one instance per class, because it allows to define pixelwise endmembers which can be derived from the representatives of each material, and interpreted geometrically.

\begin{figure}
\begin{minipage}{0.49\linewidth}
\centering
\includegraphics[scale=0.14]{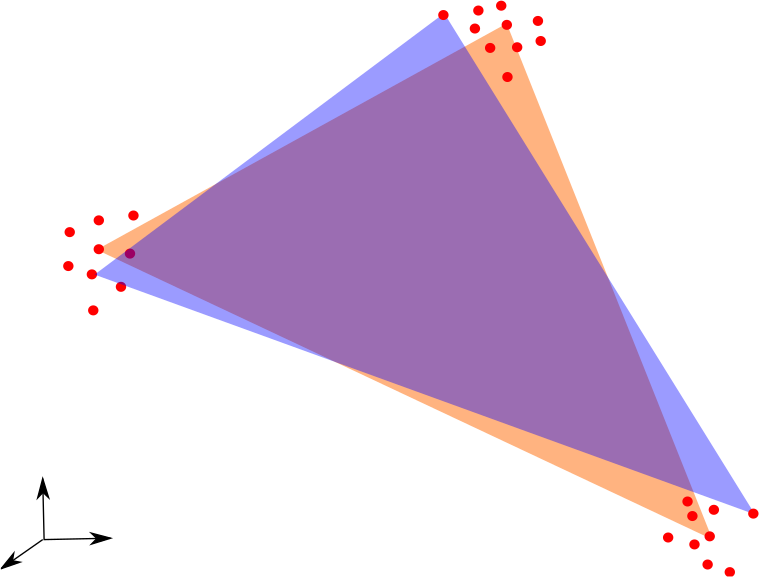}
\centerline{(a)}\medskip
\end{minipage}
\begin{minipage}{0.49\linewidth}
\centering
\includegraphics[scale=0.15]{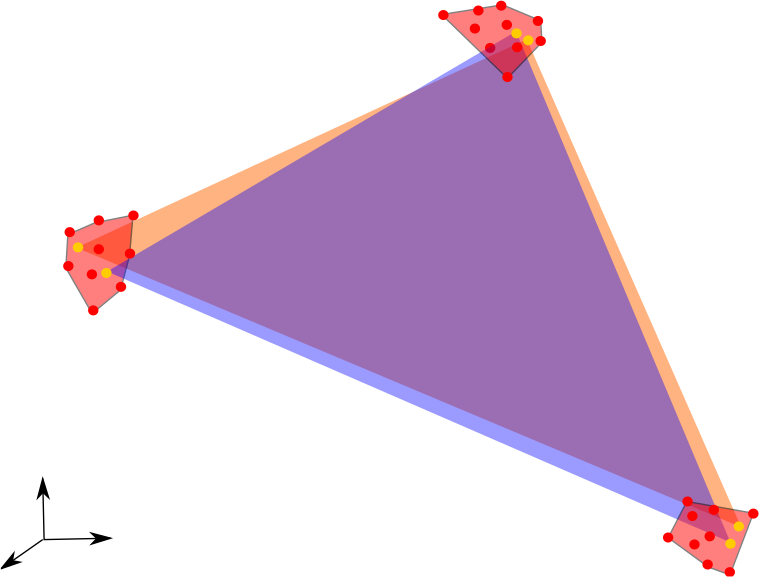}
\centerline{(b)}\medskip
\end{minipage}
\caption{Geometric interpretations of spectral unmixing with bundles. Two simplices are represented, each explaining a different pixel. (a) How endmember bundles are classically represented and (b) geometric interpretation of using FCLSU on the whole extracted dictionary. The red polytopes are the convex hull of the different bundles. The yellow points are accessible endmembers when using FCLSU, whereas they were not extracted by the EEA.}
\label{bundles_fclsu}
\end{figure}

Indeed, we can write the LMM in the context of FCLSU \cc{or any algorithm estimating one abundance coefficient per atom of the full dictionary} with spectral bundles in one pixel (we drop the pixel index here to keep the notation uncluttered) in two different ways:
\begin{equation}
\mathbf{x} = \sum_{m=1}^{Q} a_{m}\mathbf{b}_{m} = \sum_{p = 1}^{P} \left( \sum_{i = 1}^{m_{G_{p}}} {a}_{G_{p},i} \mathbf{b}_{G_{p},i} \right),
\label{bundle_decomposition}
\end{equation}
where $\mathbf{b}_{m}$ is the $m^{\textrm{th}}$ column of this dictionary, and ${a}_{G_{p},i}$ is the abundance coefficient associated to $\mathbf{b}_{G_{p},i}$. Now, if we want the global abundance of material $i$ to be $a_{p} = \sum_{i=1}^{m_{G_{p}}} {a}_{G_{p},i}$, then we have to rewrite Eq.~(\ref{bundle_decomposition}) as:
\begin{equation}
\mathbf{x} = \sum_{p = 1}^{P}  \left( \sum_{i = 1}^{m_{G_{p}}} a_{G_{p},i} \right) \left( \frac{\sum_{i = 1}^{m_{G_{p}}} {a}_{G_{p},i} \mathbf{b}_{G_{p},i}}{\sum_{i = 1}^{m_{G_{p}}} a_{G_{p},i}} \right) = \sum_{p = 1}^{P}  a_{p}  \mathbf{s}_{p},
\label{bundle_equation}
\end{equation}
with $a_{p} \triangleq  \sum_{i = 1}^{m_{G_{p}}} a_{G_{p},i} $ the total abundance coefficient for material $p$ in the considered pixel, and $\mathbf{s}_{p} \triangleq \frac{\sum_{i = 1}^{m_{G_{p}}} {a}_{G_{p},i} \mathbf{b}_{G_{p},i}}{\sum_{i = 1}^{m_{G_{p}}} a_{G_{p},i}}$. This vector actually contains a new ``equivalent" endmember for this pixel and material $p$, associated with the global ``intuitive" abundance coefficients. As a matter of fact, this new endmember is a weighted mean of all the available instances of this material, where the weights are the abundances extracted by the unmixing algorithm. Therefore, the normalized coefficients of this weighted mean are nonnegative and sum to one. This means that $\mathbf{s}_{p}$ is a convex combination of the instances of the corresponding endmember. Geometrically, each equivalent endmember belongs to the convex hull of the elements of the bundle for this material. In this sense, finding abundances with FCLSU (for example) rather than MESMA allows more freedom in terms of spectral variability: the latter constrains each pixel to come from a combination of the extracted sources, while the former theoretically allows any point inside the convex hull of the each bundle to be a local endmember. \cc{This formulation allows one to discover new endmembers as convex combinations of signatures within a class in a data-driven way}. This geometric interpretation is shown in Fig.~\ref{bundles_fclsu}. The per-pixel equivalent endmember $\mathbf{s}_{p}$ is of course only defined if at least one instance of group $G_{p}$ is active in this pixel. Otherwise, it makes no sense trying to extract spectral variability in a pixel from a material which is not present.
Similarly to what is done in the semi-supervised sparse unmixing paradigm, it makes sense to assume that only a few atoms of the bundle dictionary are going to be active in each pixel. Hence, one can add a regularization term to the formulation~(\ref{FCLSU}) so as to enforce sparse abundance vectors:
\begin{align*}
& \underset{\mathbf{A\geq \mathbf{0}}}{\textrm{arg min}} \  \frac{1}{2}||\mathbf{X}-\mathbf{B}\mathbf{A}||_{F}^{2} + \lambda ||\mathbf{A}||_{1}
\numberthis
\label{sparsity}
\end{align*}
where $||\cdot||_{1}$ denotes the sparsity-inducing $\mathcal{L}_{1}$ norm of the \cc{vectorized} matrix, i.e. the sum of the absolute values of all its entries, and $\lambda$ is a regularization term, weighting the sparsity related term w.r.t. the data fit term. Note that the ASC has been dropped here; we explain below why.

\cc{The regular $\mathcal{L}_{1}$ regularization} approach is useful to induce sparsity in the estimated abundances, but lacks two important features:
\begin{itemize}
\item The compatibility with the sum-to-one constraint. Indeed, let us assume that the ASC is enforced in problem~(\ref{sparsity}). Since the abundances are constrained to be nonnegative, a straightforward computation shows that $||\mathbf{A}||_{1} = N$. The $\mathcal{L}_{1}$ norm of the \cc{vectorized} abundance matrix is then constant, and then minimizing this quantity will result in breaking the constraint. This phenomenon is known and has been reported for instance in~\cite{6776522}, where the nonconvex $\mathcal{L}_{\frac{1}{2}}$ quasinorm instead is used to induce sparsity in the solutions. \cc{A variant of the usual $\mathcal{L}_{1}$ regularization which is compatible with the ASC is the so called collaborative sparsity regularization~\cite{Iordache2014} (described hereafter), where a mixed $\mathcal{L}_{2,1}$ mixed norm on the whole abundance matrix  to null the abundance values of some atoms of the dictionary for all the support of the image simultaneously.}
\item The structure of the bundles is not taken into account during the unmixing. Each abundance coefficient is treated independently w.r.t. sparsity. The first reason why sparsity was brought to hyperspectral image unmixing is the assumption that a few \emph{materials}, rather than \emph{atoms} of a dictionary, are active in each pixel. \cc{Neither $\mathcal{L}_{1}$ nor collaborative sparsity are designed to take the group structure into account.}
\end{itemize}
\section{Proposed abundance estimation method}
\label{proposed}
\subsection{Group sparsity inducing mixed norms}
\begin{figure}
\centering
\rotatebox{-90}{\includegraphics[scale=1.8]{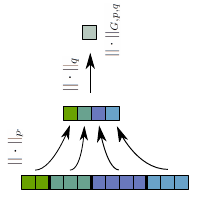}}
\caption{Illustration on how the  $\mathcal{L}_{G,p,q}$ norm operates on a vector, given the group structure $G$.}
\label{pqnorm}
\end{figure}

All the penalties we are going to detail in the following are based on applying a mixed norm on the abundance vector (in each pixel), which is endowed with a group structure $G$, partitioning the $Q = mP$ endmembers extracted by the $n$ runs of VCA on random subsets into $P$ groups (as many as the number of materials to unmix). We drop the pixel index for simplicity of the notation. In the most general form, the group two-level mixed $\mathcal{L}_{G,p,q}$ norm is defined, for any two positive real numbers $p$ and $q$ as~\cite{kowalski2013social}:
\begin{equation}\label{eq_col_social}
||\mathbf{a}||_{G,p,q} \triangleq \left(\sum_{i=1}^P \left( \sum_{j=1}^{m_{G_{i}}} {|a_{{G_i},j}|}^p \right) ^\frac{q}{p}\right)^\frac{1}{q} = \left(\sum_{i=1}^P ||\mathbf{a}_{G_i}||_p^q\right)^\frac{1}{q},
\end{equation}
where $\mathbf{a}_{G_i}$ is a subvector of the abundance vector ($\mathbf{a}$) comprising all the abundance coefficients associated to the \cc{indices (in the most natural case, the endmember candidates corresponding to the same material)} of group $G_{i}$. This equation only defines a true norm for $p,q \geq 1$ (and also for $p$ or $q = \infty$, by taking limits). As explained in Fig.~\ref{pqnorm}, the idea is to take the $p$ norm of each of the $P$ subvectors of coefficients defined by the groups, to store the results in a $P$-dimensional vector, for which we are going to compute the $q$ norm.
We will see that with smart choices of $p$ and $q$, different types of sparsity can be obtained when this mixed norm is used as a regularizer in the unmixing with bundles. 

\begin{figure}
\begin{center}
\rotatebox{0}{\includegraphics[scale=1.1]{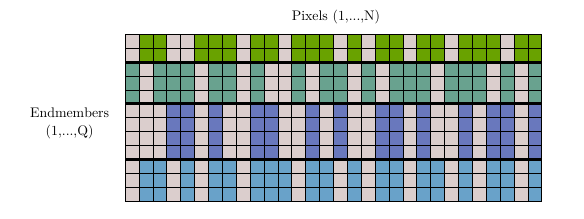}}
\caption{Effect of the group LASSO penalty on the abundance matrix. The group structure is shown in colors (the rows of the matrix have been sorted for more clarity). Inactive entries of the matrix are in gray. A small number of groups is selected in each pixel, but within each group the matrix is dense.}
\label{group_lasso}
\end{center}
\end{figure}

The definition of the  $\mathcal{L}_{G,p,q}$ norm can easily be extended to a matrix $\mathbf{A} \in \mathbb{R}^{Q\times N}$, using the same expression, operating columnwise, and summing the results on all pixels: 
\begin{equation}
||\mathbf{A}||_{G,p,q} \triangleq \sum_{k=1}^{N} ||\mathbf{a}_{k}||_{G,p,q}.
\end{equation}
Here, we are interested in norms which can handle any group structure, while enforcing several types of sparsity. For instance, the use of sparsity in SU is based on the assumption that a few materials are active in each pixel. If the dictionary of endmembers has a group structure, it makes sense to enforce sparsity on the number of groups, rather than on the total number of signatures. This rationale is the basis of the so-called group LASSO~\cite{meier2008group}. This method uses the  $\mathcal{L}_{G,2,1}$ norm, which enforces sparsity on the vector whose entries are the $||\mathbf{a}_{G_i}||_{2}$. This means that when one of these entries is zero, the whole group is discarded entirely since the vector $\mathbf{a}_{G_i}$ has a zero norm. Within each group, there is no sparsity and thus most or all signatures are likely to be active. The effect of this penalty on a matrix $\mathbf{A} \in \mathbb{R}^{Q\times N}$ is shown in Fig.~\ref{group_lasso}.
\cc{Note that if one defines a group structure $H$, given by $P$ groups of coefficients corresponding to the $N$ abundance values of each atom of the dictionary on the whole support of the image, and stores them in a vector $\textrm{vec}({\mathbf{A}})\in\mathbb{R}^{NP}$ (by stacking the columns), then the group LASSO with this group structure is equivalent to the collaborative LASSO of~\cite{Iordache2014}. The optimization problem solved here is
\begin{align*}
& \underset{\mathbf{A}\geq \mathbf{0}, \ \mathds{1}_{P}^{\top}\mathbf{A} = \mathds{1}_{N}^{\top}}{\textrm{arg min}} \  \frac{1}{2}||\mathbf{X}-\mathbf{B}\mathbf{A}||_{F}^{2} + \lambda ||\textrm{vec}({\mathbf{A})}||_{H,2,1}
\numberthis
\label{collaborative_sparsity}
\end{align*}
However, in that case, the information provided by the knowledge of the bundles is not taken into account}.

\begin{figure}
\begin{center}
\rotatebox{0}{\includegraphics[scale=1.1]{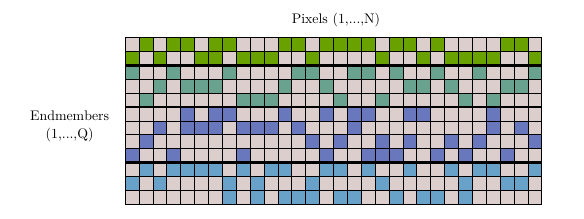}}
\caption{Effect of the elitist LASSO penalty on the abundance matrix. A small number of instance in each group is selected in each pixel, but all or almost all groups are active.}
\label{elitist_lasso}
\end{center}
\end{figure}

\begin{figure}
\begin{center}
\rotatebox{-90}{\includegraphics[scale=1.1]{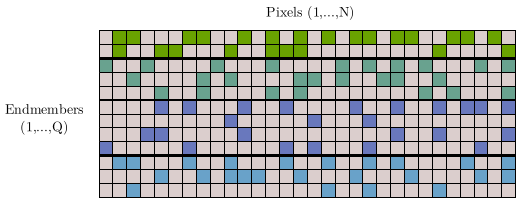}}
\caption{Effect of the fractional LASSO penalty on the abundance matrix. A small number of instance in each group is selected in each pixel, and the mixture is also sparse within each group.}
\label{fractional_lasso}
\end{center}
\end{figure}

In some cases, for example when we deal with a small number of groups, and we have reasons to believe that there is only one or a few instances of each group which are active in a pixel, we may expect within-group sparsity, without group sparsity. In this case the elitist LASSO penalty~\cite{kowalski2009sparsity} is suited to the problem, since it uses the $\mathcal{L}_{G,1,2}$ norm, which promotes a small value of the $\mathcal{L}_{1}$ norms of each $\mathbf{a}_{G_i}$. The effect of this penalty is shown in Fig.~\ref{elitist_lasso}.
Using a penalty which enforces both group sparsity and global sparsity (on the total number of active signatures) also seems appealing. A seemingly natural way to do so with the proposed framework is to take both $p = q = 1$, to enfore sparsity on both norms that are computed in sequence. However, it turns out that the $\mathcal{L}_{1,1}$ norm actually is the same as the regular $\mathcal{L}_{1}$ norm \cc{of the vectorized matrix}, i.e. using these values for $p$ and $q$ actually breaks the group structure of the coefficients, and is of course still incompatible with the ASC. The sparse group LASSO~\cite{simon2013sparse} uses a combination of the group lasso penalty and a classical $\mathcal{L}_{1}$ norm to benefit from both properties. It was recently used in a sparse SU context in \cite{iordache2011hyperspectral}. However, in this case, benefiting from both penalties comes at the cost of having two regularization parameters to tune. In our case, recall that this penalty is also still at odds with the ASC due to the presence of the $\mathcal{L}_{1}$ norm in the objective. The ASC can also be contradictory with sparsity in some other configurations: for instance, if each material has only one representative, the group LASSO reduces to the regular LASSO and the ASC conflicts with the objective. In order to avoid this issue, we are also using a fractional case, with the $\mathcal{L}_{G,1,q}$ ``norm", with $q = \frac{a}{b}$ ($a$ and $b\neq 0\in \mathbb{N}$) and $0 < \frac{a}{b} < 1$. This penalty is no longer a norm, because we lose convexity due to the fact that $q < 1$, but it has the advantage of enforcing both group sparsity and within-group sparsity in a compact formulation, without conflicting with the ASC anymore. In addition, the $\mathcal{L}_{q}$ quasinorm $q < 1$ is a better approximation of the $\mathcal{L}_{0}$ norm than the $\mathcal{L}_{1}$ norm. The effect of the  $\mathcal{L}_{G,1,q}$ penalty on the abundance matrix is shown in Fig.~\ref{fractional_lasso}. The choice of the fraction is an additional parameter, but we will see that it turns out to be relatively easy to tune.

Geometrically, the group lasso encourages a small number of materials (i.e. groups) to be active in each pixel, but tends to prefer a dense mixture within each group. Hence, the pixelwise endmembers tend to lie around the center of mass of each bundle. For the elitist penalty, a large number of groups are active in each pixel. However, within each bundle, a small number of endmember candidates will be active, which translates geometrically into the fact that the pixelwise endmbers will lie inside polytopes with a few number of vertices. The fractional penalty has the same effect as the elitist one, except that in addition only a few groups are going to be active in each pixel.

\subsection{Optimization}

With either of those penalties, the optimization problem to solve is:
\begin{equation}
\underset{\mathbf{A}}{\textrm{arg min}} \ \frac{1}{2}||\mathbf{X} - \mathbf{B}\mathbf{A}||_{F}^{2} + \lambda||\mathbf{A}||_{G,p,q} + \mathcal{I}_{\Delta_{Q}}(\mathbf{A}).
\label{sparse_bundles}
\end{equation}
where, $\mathcal{I}_{\Delta_{Q}}$ is the indicator function of the unit simplex $\Delta_{Q}$ with $Q$ vertices. \cc{It is defined, first for a vector $\mathbf{a}\in\mathbb{R}^{P}$, as:
\begin{equation}
    \mathcal{I}_{\Delta_{Q}}(\mathbf{a}) =\begin{cases}
                  0 \ \textrm{if} \ \mathbf{a}\in \Delta_{Q}\\
                   +\infty \ \textrm{otherwise}\\
              \end{cases},
\end{equation}
This can be extended to a matrix $\mathbf{A}\in \mathbb{R}^{P\times N}$ via
\begin{equation}
    \mathcal{I}_{\Delta_{Q}}(\mathbf{A}) = \sum_{k = 1}^{N} \mathcal{I}_{\Delta_{Q}}(\mathbf{a}_{\cdot k}).
\end{equation}
with $\mathbf{a}_{\cdot k}$ the $k^{\textrm{th}}$ column of $\mathbf{A}$.}\\

Note that after solving this problem, in order to obtain the global abundances, one simply has to sum the abundances within each instance of each group, as described in section~\ref{bundles}. The optimization problem~(\ref{sparse_bundles}) is convex for both the group and elitist penalties, but not for the fractional one.

In any case, since the problem is not differentiable, we are going to use the Alternating Direction Method of Multipliers (ADMM)~\cite{boyd2011} to solve it. Convergence to the global minimum is automatically guaranteed for the group and elitist penalties. For the nonconvex case, as we will see, the situation is more complex. The ADMM was designed to tackle convex problems, but it has been more and more (successfully) used for nonconvex problems as well, and recent works~\cite{wang2015global} show that if the nonconvex function satisfies some conditions, the ADMM is proven to converge to a stationary point in the nonconvex case. One of these cases includes the $\mathcal{L}_{p}$ quasinorm for $p <1$. This results remains to be shown in the mixed $\mathcal{L}_{G,1,q}$ norm with $q<1$ case, (but it is likely to satisfy the same conditions, being ``less nonconvex" than the $\mathcal{L}_{p}$ quasinorm, because the unit ball of such a norm will have some nonconvex facets, but not all since some of them will be similar to the facets of the $\mathcal{L}_{1}$ ball).

The next two sections lay out the algorithm to solve the optimization problem with the group and elitist penalties, while the last one shows how to handle the fractional case. 

\subsubsection{Group and elitist cases}

The optimization problem to solve in both cases is convex, and can be handled in the same fashion. In order to use the ADMM, we need to introduce auxiliary split variables, which will decompose problem~(\ref{sparse_bundles}) into easier subproblems. Let us then rewrite this problem as:
\begin{align*}
& \underset{\mathbf{A}}{\textrm{arg min}} \  \frac{1}{2}||\mathbf{X} - \mathbf{B} \mathbf{A}||_{F}^{2} + \lambda ||\mathbf{U}||_{G,p,q} + \mathcal{I}_{\Delta_{Q}}(\mathbf{V}) \\
& \textrm{s.t.} \ \mathbf{U} = \mathbf{A}, \ \mathbf{V}=\mathbf{A}.
\numberthis
\label{collaborative_optim}
\end{align*}
The ADMM  consists in expressing the constrained problem defined in Eq.~(\ref{collaborative_optim}) in an unconstrained  way using an Augmented Lagrangian (AL), and then minimizing it iteratively and alternatively for each of the variables introduced, before finally updating the Lagrange Multipliers appearing in the AL (the so-called \emph{dual update}). $\rho$ is the \emph{barrier} parameter weighting the AL terms (which we will set to $\rho = 10$ for all the algorithms). Here, the Augmented Lagrangian writes:
\begin{align*}
& \mathcal{L}(\mathbf{A},\mathbf{U},\mathbf{V}) = \  \frac{1}{2}||\mathbf{X} - \mathbf{B} \mathbf{A}||_{F}^{2} + \lambda ||\mathbf{U}||_{G,p,q} + \mathcal{I}_{\Delta_{Q}}(\mathbf{V}) \\
& + \frac{\rho}{2}  ||\mathbf{A} - \mathbf{U} - \mathbf{C}||_{F}^{2} + \frac{\rho}{2} ||\mathbf{A} - \mathbf{V} - \mathbf{D}||_{F}^{2}
\numberthis
\end{align*}
where $\mathbf{C}$ and $\mathbf{D}$ are the set of dual variables. 
Each of the updates is supposed to be straightforward. The update w.r.t. $\mathbf{A}$ only involves quadratic terms. and the updates w.r.t. $\mathbf{U}$ and $\mathbf{V}$ can be readily obtained, provided the proximal operators~\cite{combettes2011proximal} of the mixed norm, and of the indicator function of the simplex can be easily computed. The latter can indeed be easily obtained with the algorithm of~\cite{condat2014fast}. For both the group and elitist penalties, the proximal operators have closed forms, which we provide now.

The proximal operator of the group penalty is simply a group version of the so-called \emph{block} soft thresholding operator, i.e. the proximal operator of the $\mathcal{L}_{2}$ norm:
\begin{equation}
\textbf{prox}_{\tau ||\cdot||_{G,2,1}}(\mathbf{v}) = \left[ 
\begin{array}{c}
\textbf{soft}_{\tau}(\mathbf{v}_{G_{1}}) \\
\vdots \\
\textbf{soft}_{\tau}(\mathbf{v}_{G_{P}})
\end{array} \right].
\end{equation}
The block soft thresholding is denoted by $\textbf{soft}_{\tau}$, where $\tau$ is the scale parameter of this operator:
\begin{equation}
\forall \mathbf{u}\in\mathbb{R}^{P},\textbf{soft}_{\tau}(\mathbf{u}) = \left( 1-\frac{\tau}{||\mathbf{u}||_{2}} \right)_{+}\mathbf{u},
\label{block_soft}
\end{equation}
where $\cdot_{+} = \textrm{max}(\cdot,0)$ (and we have $\textbf{soft}_{\tau}(\mathbf{0}) = 0$).

The proximal operator for the elitist norm is a bit more complex, but has a closed form (derived in~\cite{kowalski2009sparse}), which involves the \emph{regular} soft thresholding operator, the proximal operator of the $\mathcal{L}_{1}$ norm. We recall that
\begin{equation}
\textrm{soft}_{\tau}(\mathbf{u})_{i} = \textrm{sign}(u_{i})(|u_{i}| -\tau)_{+}.
\end{equation}
With this definition, the proximal operator of the elitist penalty is given as:
\begin{equation}
\textbf{prox}_{\tau ||\cdot||_{G,1,2}}(\mathbf{v}) =  \left[ 
\begin{array}{c}
\textrm{soft}_{\gamma_{1}}(\mathbf{v}_{G_{1}}) \\
\vdots \\
\textrm{soft}_{\gamma_{P}}(\mathbf{v}_{G_{P}})
\end{array} \right],
\end{equation}
where the soft thresholding is applied entrywise, and 
\begin{equation}
\gamma_{i} = \frac{\tau}{1+\tau} ||\mathbf{v}_{G_{i}}||_{1}, \ \forall i \in  [\![1,P]\!].
\end{equation}
The ADMM procedure for any case is summarized in Algorithm~\ref{ADMM_group}, in which $\textbf{prox}_{\cdot}$ denotes the proximal operator of the function in index, and $\textbf{proj}_{\Delta_{Q}}$ denotes the proximal operator of the indicator function of the simplex, a simple projection. $\mathbf{I}_{\cdot}$ denotes the identity matrix whose size is in index (we only provide one dimension for brevity since the matrix is square). Note that the proximal operators are carried out columnwise.
\begin{algorithm}
 \KwData{$\mathbf{X}$, $\mathbf{B}$}
 \KwResult{$\mathbf{A}$}
 Initialize $\mathbf{A}$ and choose $\lambda$ \;
 \While{ADMM termination criterion is not satisfied}{
  $\mathbf{A} \leftarrow (\mathbf{B}^{\top} \mathbf{B} + 2 \rho \mathbf{I}_{Q})^{-1} (\mathbf{B}^{\top} \mathbf{X}+ \rho(\mathbf{U} + \mathbf{V} + \mathbf{C} + \mathbf{D}))$\
  $\mathbf{U} \leftarrow \textbf{prox}_{(\lambda/\rho ) ||\cdot||_{G,p,q}}(\mathbf{A} - \mathbf{C})$ \;
  $\mathbf{V} \leftarrow  \textbf{prox}_{\mathcal{I}_{\Delta_{Q}}}(\mathbf{A} - \mathbf{D}) = \textbf{proj}_{\Delta_{Q}}(\mathbf{A} - \mathbf{D})$ \;
  $\mathbf{C} \leftarrow \mathbf{C} + \mathbf{U} - \mathbf{A}$ \; 
 $\mathbf{D} \leftarrow \mathbf{D} + \mathbf{V} - \mathbf{A}$ \; }
 \caption{ADMM to solve problem~(\ref{collaborative_optim}) in the convex cases).}
 \label{ADMM_group}
\end{algorithm}

\subsubsection{Fractional case}

The problem is more complex for the fractional mixed norm. As we have pointed out above, there is no proof that the mixed $\mathcal{L}_{G,1,q}$ norm with $q<1$ satisfies the required properties for the ADMM to converge to a local minimum. However, in our problem, with an appropriate variable splitting scheme, we can express the fractional case for problem~(\ref{sparse_bundles}) as a $\mathcal{L}_{q}$ regularized constrained least squares problem.

Let us suppose for simplicity (and without loss of generality), that the rows of $\mathbf{A}$ and the columns of $\mathbf{B}$ have been sorted so that, in each pixel, the abundance vector has the following form:
\begin{align*}
\mathbf{a} =  [a_{1,1}, a_{1,2}, \cdots, a_{1,m_{G_1}},& a_{2,1}, a_{2,2}, \cdots, a_{2,m_{G_2}},\cdots ,\\
& a_{G_{P},1}, a_{G_{P},2}, \cdots, a_{G_{P},m_{G_P}}] ^{\top}.
\end{align*}
Recall that $m_{G_{i}}$ is the number of instances of one of the $P$ groups, in this case $G_{i}$.\\
The problem we want to solve is:
\begin{equation}
\underset{\mathbf{A}}{\textrm{arg min}} \  \frac{1}{2} ||\mathbf{X}-\mathbf{BA}||_{F}^{2} + \lambda ||\mathbf{A}||_{G,1,q}^{q} + \mathcal{I}_{\Delta_{Q}}(\mathbf{A}),
\label{pb}
\end{equation}
with $Q =  \sum_{i=1}^{P} m_{G_{i}}$ the total number of signatures in the bundle dictionary $\mathbf{B} \in \mathbb{R}^{L\times Q}$. With the following variable splitting scheme, Problem .~(\ref{pb}) is equivalent to:
\begin{align*}
& \underset{\mathbf{A}}{\textrm{arg min}} \ \frac{1}{2}||\mathbf{X}-\mathbf{BA}||_{F}^{2} + \lambda ||\mathbf{U}||_{q}^{q} + \mathcal{I}_{\Delta_{Q}}(\mathbf{V}) \\
& \textrm{s.t.} \ \mathbf{MA} = \mathbf{U} \in \mathbb{R}^{P \times N},\  \mathbf{A} = \mathbf{V}
\numberthis
\end{align*}
with 
\begin{equation} \mathbf{M} = \left[ \begin{array}{c c c c c c c c c c c c}
1& \cdots& 1 & 0 & \cdots & 0  & 0& \cdots &0  & 0 &\cdots  &  0 \\
0 & \cdots & 0 & 1 & \cdots & 1 & 0 & \cdots & 0 & 0 &\cdots  & 0 \\
\vdots & & \vdots & \vdots & &  \vdots &\vdots & &\vdots & & &\vdots\\
0  & \cdots & 0  & 0 & \cdots & 0 &0 & \cdots & 0  & 1 & \cdots & 1 \\
\end{array} \right],
\end{equation}
where $\mathbf{M}\in \mathbb{R}^{P\times Q}$, and the  $i^{\textrm{th}}$ row having $m_{G_i}$ consecutive ones.

This way, we have reduced the optimization problem to a $\mathcal{L}_{q}$ regularized least squares problem, where the variable on which the fractional norm is applied is a vector whose entries are the $\mathcal{L}_{1}$ norms of the abundance coefficients in each group. Note that the new problem is equivalent to the original one only thanks to the nonnegativity constraint, which allowed us to turn the $\mathcal{L}_{1}$ norm into linear constraints. This way, the convergence of the ADMM for our nonconvex problem is, in theory, guaranteed~\cite{wang2015global}, should we be able to compute exact updates for all the subproblems of the ADMM.

However, even after this simplification of the problem, an issue remains: there is no closed form expression or known algorithm (to the best of our knowledge) to compute exactly the shrinkage operator of the $\mathcal{L}_{q}$ quasinorm (to the power $q$) when $q < 1$, except when $q=\frac{1}{2}$ or $q= \frac{2}{3}$~\cite{cao2013fast}. Here, we prefer the term ``shrinkage operator" to the term ``proximal operator" because the latter is usually defined for convex functions only. In addition, this operator is a discontinuous function, because of the nonconvexity of the quasinorm~\cite{yukawa2013lp}. This limits the applicability of proximal methods to solve this type of problems, and other types of algorithms (or of nonconvex sparsity inducing penalties) have been investigated in remote sensing (see e.g.~\cite{tuia2016nonconvex} and references therein).

In our case, in order to be able to apply ADMM nonetheless, we need an explicit shrinkage operator. We resort to an approximate $q$-shrinkage operator $S_{q,\tau}$, as defined in~\cite{woodworth2016compressed}:
\cc{
\begin{equation}
\forall \mathbf{u} \in \mathbb{R}^{P}, \ {S_{q,\tau}}(\mathbf{u})_{i} = \textrm{sign}(u_{i})(|u_{i}| - \tau^{2-q}|u_{i}|^{q-1})_{+}.
\label{pshrinkage}
\end{equation}}
This operator reduces to the soft thresholding operator when $q =1$ and to the \emph{hard thresholding} operator when $q = 0$. The hard thresholding operator is closely related to the shrinkage operator of the $\mathcal{L}_{0}$ norm~\cite{woodworth2016compressed}. In addition, it can be shown (\cite{woodworth2016compressed}, theorem II.4) that the operator of Eq.~(\ref{pshrinkage}) is actually the exact shrinkage operator of a nonconvex function (which we will denote as $f_{q}$) with desirable properties: it is separable w.r.t. each entry of $\mathbf{x}$, with $f_{q}(\mathbf{x}) = \sum_{i = 1}^{P} g_{q}(x_{i})$, and the function $g_{q}$ is even, continuous, strictly increasing and concave for $x_{i}>0$, differentiable everywhere except in 0, and satisfies the triangle inequality. This function behaves in a way similar to the absolute value for small values of its argument, and more like the absolute value taken to the power $q$ for larger arguments (see~\cite{woodworth2016compressed} for a graphical representation). For $q =1$, this penalty function is the absolute value, but in general, unfortunately, there is no analytical expression for it. This result is interesting because we have an explicit shrinkage operator with nice properties to use with any proximal algorithm, to the cost of having a regularizer without an explicit expression. Nevertheless, the convergence of the ADMM in the nonconvex case remains to be proven, since we replaced the $\mathcal{L}_{q}$ norm with another nonconvex penalty, which should itself satisfy the required properties of~\cite{wang2015global} in order to guarantee convergence.

Finally, the optimization problem we solve is:
\begin{align*}
& \underset{\mathbf{A}}{\textrm{arg min}} \ \frac{1}{2}||\mathbf{X}-\mathbf{BA}||_{F}^{2} + \lambda f_{q}(\mathbf{U}) + \mathcal{I}_{\Delta_{Q}}(\mathbf{V}) \\
& \textrm{s.t.} \ \mathbf{MA} = \mathbf{U}, \  \mathbf{A} = \mathbf{V}.
\numberthis
\label{optim_fractional}
\end{align*}
The AL writes:
\begin{align*}
\mathcal{L}(\mathbf{A},\mathbf{U},\mathbf{V}) & = \frac{1}{2}||\mathbf{X}-\mathbf{BA}||_{F}^{2} + \lambda f_{q}(\mathbf{U}) + \mathcal{I}_{\Delta_{Q}}(\mathbf{V}) \\
& + \frac{\rho}{2}||\mathbf{MA}-\mathbf{U} - \mathbf{C}||_{F}^{2} + \frac{\rho}{2}||\mathbf{A}-\mathbf{V}-\mathbf{D}||_{F}^{2}
\numberthis
\end{align*}
and we can use the ADMM algorithm to minimize it. The optimization procedure is described in Algorithm~\ref{fractional_sparsity_ADMM} (the approximate $q$-shrinkage is performed coordinate-wise).
\begin{algorithm}
 \KwData{$\mathbf{X}$, $\mathbf{B}$}
 \KwResult{$\mathbf{A}$}
 Initialize $\mathbf{A}$ and choose $\lambda$ \;
 \While{ADMM termination criterion is not satisfied}{
  $\mathbf{A} \leftarrow (\mathbf{B}^{\top} \mathbf{B} + \rho \mathbf{M}^{\top} \mathbf{M}+ \rho \mathbf{I}_{Q})^{-1} (\mathbf{B}^{\top} \mathbf{X} + \rho \mathbf{M}^{\top}(\mathbf{U} + \mathbf{V}) + \rho (\mathbf{C} + \mathbf{D}))  $ \;
  $\mathbf{U} \leftarrow S_{q,\lambda/\rho}(\mathbf{MA} - \mathbf{C})$ \;
  $\mathbf{V} \leftarrow  \textbf{prox}_{\mathcal{I}_{\Delta_{P}}}(\mathbf{A} - \mathbf{D}) = \textbf{proj}_{\Delta_{P}}(\mathbf{A} - \mathbf{D})$ \;
  $\mathbf{C} \leftarrow \mathbf{C} + \mathbf{U} - \mathbf{M}\mathbf{A}$ \; 
 $\mathbf{D} \leftarrow \mathbf{D} + \mathbf{V} - \mathbf{A}$ \; }
 \caption{ADMM to solve problem~(\ref{optim_fractional}).}
 \label{fractional_sparsity_ADMM}
\end{algorithm}
\subsection{Computational complexity}
For the three proposed algorithms, the most costly operations involved are matrix products and solving linear systems. Hence the computational complexity of the algorithms for the group and elitist penalties is $\mathcal{O}(Q^2(L+N))$ (assuming $Q \ll N$). The complexity for FCLSU is the same but a much faster convergence is expected since the constraints are less complex and much less iterations will be required to reach convergence. Since the fractional penalty involves additional matrix products using matrix $\mathbf{M}$, then the complexity for the corresponding algorithm is slightly higher and is $\mathcal{O}(Q^2(L+N+P) + QPN)$. A significant increase in running time is then expected for the fractional penalty when the number of endmembers $P$ gets large.
\section{Results}
\label{results}
\subsection{Synthetic data}
%\subsubsection{Semisupervised case}
First, we test the three proposed penalties in a semi-supervised scenario, where the possible instances of each endmember are known beforehand. This case is similar to the now well-known and widespread sparse unmixing approaches~\cite{Iordache2011,7293641}. The main difference with respect to regular sparse unmixing is that here the spectral library incorporates several variants of each endmember in order to account for the variability of the scene. The group structure $G$ is then known a priori as well. The objective of this dataset is to compare the performance of the proposed penalties in terms of abundance estimation and variability retrieval, i.e. the core of the contribution of this paper, when the bundle extraction is assumed to have been carried out efficiently.

In order to show the interest of the proposed penalties, we designed a synthetic dataset corresponding to a challenging unmixing scenario, with 20 different materials present in the scene. We then selected randomly $P=20$ signatures from the United States Geological Survey (USGS) spectral library~\cite{kokaly2017usgs}, comprising 224 spectral bands in the visible and near infra-red parts of the electromagnetic spectrum. For each of those 20 signatures, spectral variants were generated by modifiying the spectra using several effects: scaling variations, accounting for local changes in illumination conditions~\cite{drumetztip}, a nonlinear quadratic (in the original signature) perturbation to mimick intrinsic variability effects, and i.i.d white Gaussian noise. 20 such variants were randomly generated for each material, providing a dictionary of $Q = 400$ spectral signatures, divided into $P=20$ groups. In each pixel, only one signature is active for each endmember. \ccc{We focus on determining whether the group information is beneficial to distinguish between very correlated materials.}
 $50\times 50$ sparse (between 1 and 3 active materials in each pixel) abundance maps were computed using Gaussian Random Fields, so as to give them a spatial coherence (which is not used by the algorithm but is helpful to visualize the results).
 
We tested the performance of \cc{6} algorithms on this dataset: the FCLSU algorithm, \cc{classical sparse unmixing using $\mathcal{L}_{1}$ regularization~\cite{Iordache2011}, \cc{the collaborative sparsity approach of~\cite{Iordache2014}} (these three algorithms do not take the group structure into account)} and the three proposed penalties: group penalty, elitist penalty, and fractional penalty. In each case, we optimize the regularization parameter via a grid search to obtain the best abundance estimation performance ($RMSE(\hat{\mathbf{A}})$, see below). The FCLSU algorithm was implemented using ADMM as well, to allow a fair comparison for the running times of the various algorithms. We ran all the algorithms for 1000 iterations or stopped them when the relative variations (in norms) of the abundance matrix went below $10^{-6}$. For the fractional penalty, the grid search is also carried out w.r.t. the choice of the fraction. \cc{We tested values in three orders of magnitude, between $10^{-3}$ and $10^{-2}$, $10^{-2}$ and $10^{-1}$, and $10^{-1}$ and $0.9$, with 9 equally spaced steps for each}. To quantitatively assess the abundance estimation performance, we use the abundance Root Mean Squared Error $RMSE(\hat{\mathbf{A}})$ between the estimated abundances $\hat{\mathbf{a}_{k}}$ for each material (summing the abundances of all the representatives for each group) and the true ones $\mathbf{a}_{k}$ in each pixel:
\begin{equation}
\vspace{-0.135cm}
RMSE(\hat{\mathbf{A}}) = \frac{1}{N} \sum_{k=1}^{N} \sqrt{ \frac{1}{P}\sum_{p=1}^{P} (a_{pk}-\hat{a}_{pk})^{2}}
\end{equation}
where the abundance matrix comprises only the global abundance coefficients.

To assess the precision of the abundance estimation within each group, we also use this metric for each individual atom of the dictionary, which we call in that case $RMSE_{G}(\hat{\mathbf{A}})$:
\begin{equation}
\vspace{-0.135cm}
RMSE_{G}(\hat{\mathbf{A}}) = \frac{1}{N} \sum_{k=1}^{N} \sqrt{ \frac{1}{P}\sum_{p=1}^{P\times m_G} (a_{pk}-\hat{a}_{pk})^{2}},
\end{equation}
where $m_G = 20$ is the number of instances in each group, and here the abundance matrix used is the full bundle dictionary.

To compare the performance in terms of material variability retrieval, we also use the Root Mean Squared Error between the true signatures and the estimated ones:\cc{
\begin{equation}
RMSE(\hat{\mathbf{S}}) = \frac{1}{NP} \sum_{k=1}^{N} \sum_{p=1}^{P} \frac{1}{\sqrt{L}} ||\mathbf{b}_{pk}-\hat{\mathbf{s}}_{pk}||_{2}
\end{equation}}
and the overall spectral angle mapper (SAM) between the true signature $\mathbf{b}_{pk}$ used for material $p$ in pixel $k$ and the estimated ones $\hat{\mathbf{s}}_{pk}$ (using Eq.~\eqref{bundle_equation}):
\begin{equation}
SAM = \frac{1}{NP} \sum_{k=1}^{N} \sum_{p=1}^{P} \textrm{acos}\left(\frac{\mathbf{b}_{pk}^{\top}{\hat {\mathbf{s}}_{pk}}}{||\mathbf{b}_{pk}||_{2}||\hat{\mathbf{s}}_{pk}||_{2}} \right)
\end{equation}
All the quantitative results \cc{for the all algorithms, except $\mathcal{L}_{1}$ sparsity and collaborative sparsity (we explain below why)} are gathered in Table~\ref{table_synth}.
\begin{table}
\begin{center}
\scalebox{0.7}{
\begin{tabular}{| c | c | c | c | c|}
  \hline 
  \backslashbox{Metric}{Algorithm}   &  FCLSU & Group     & Elitist  & Fractional \\ 
  \hline
 $RMSE(\hat{\mathbf{A}})$ & 0.0103 & \textcolor{blue}{0.0072}  & 0.0158 & \textcolor{red}{0.0064}   \\
  \hline
 $RMSE_{G}(\hat{\mathbf{A}})$ &   \textcolor{red}{0.0166}     &   0.0183 & \textcolor{blue}{0.0169} & 0.0209    \\

   \hline
 $RMSE(\mathbf{\hat{S}})$ & 0.0342   & \textcolor{blue}{0.0339}  & 0.0382   & \textcolor{red}{0.0337}  \\
   \hline
 SAM (degrees) & 2.026   & \textcolor{blue}{1.749}  & 2.148   & \textcolor{red}{1.738}  \\
    \hline
 Running time (s) & 16   & 73  & 74  & 95  \\
  \hline
 $\lambda$  & $\times$   & 0.006  & 0.01   & 0.02  \\ \hline
\end{tabular}}
\end{center}
\caption{Different metrics for each tested algorithm on the synthetic data. The best value is red, and the second best is in blue. The regularization parameters are also reported , when applicable.}
\label{table_synth}
\end{table}
\cc{Before describing these results, let us comment on the performance of the $\mathcal{L}_{1}$ and collaborative sparsity. In both cases, we have performed a grid search on the value of $\lambda$ to obtain the best performance in terms of $RMSE(\hat{\mathbf{A}})$. As shown in Fig.~\ref{sparse_grid_search}, in both cases, the abundance estimation results on the synthetic data are optimal (and almost equal to those of FCLSU) for low values of the regularization parameters, which means that in that case the sparsity penalty is actually detrimental to the unmixing performance, since the algorithms prefers a configuration without sparsity. One of the reasons we have identified for this behavior is that some of the randomly chosen endmembers are very correlated (low brightness signatures, see Fig.~\ref{bundles_synth}). In this case, if the group information is not exploited during the unmixing, these materials are not unmixed, and the algorithm prefers to put the global abundance maps of several of them to zero, and to put all the abundance weight on the rest. This proves that using the group information, if available, is crucial for the unmixing performance of correlated endmembers. We can also see from this figure that the regular $\mathcal{L}_{1}$ sparsity performs slightly worse than collaborative sparsity for small regularization parameters. This could be due to the fact that the ASC is not enforced when using classical sparsity, resulting in a slight loss of accuracy in the abundance estimation. For larger regularization parameters, on the contrary standard sparsity is more adapted since all the materials can be randomly selected to be active in each pixel. In any case, the abundance maps and matrices of the collaborative sparsity approach will still be shown in the following figures, for visual comparison with the competing algorithms, with $\lambda = 0.1$.}
\cc{\begin{figure}
\begin{center}
\includegraphics[scale=0.225]{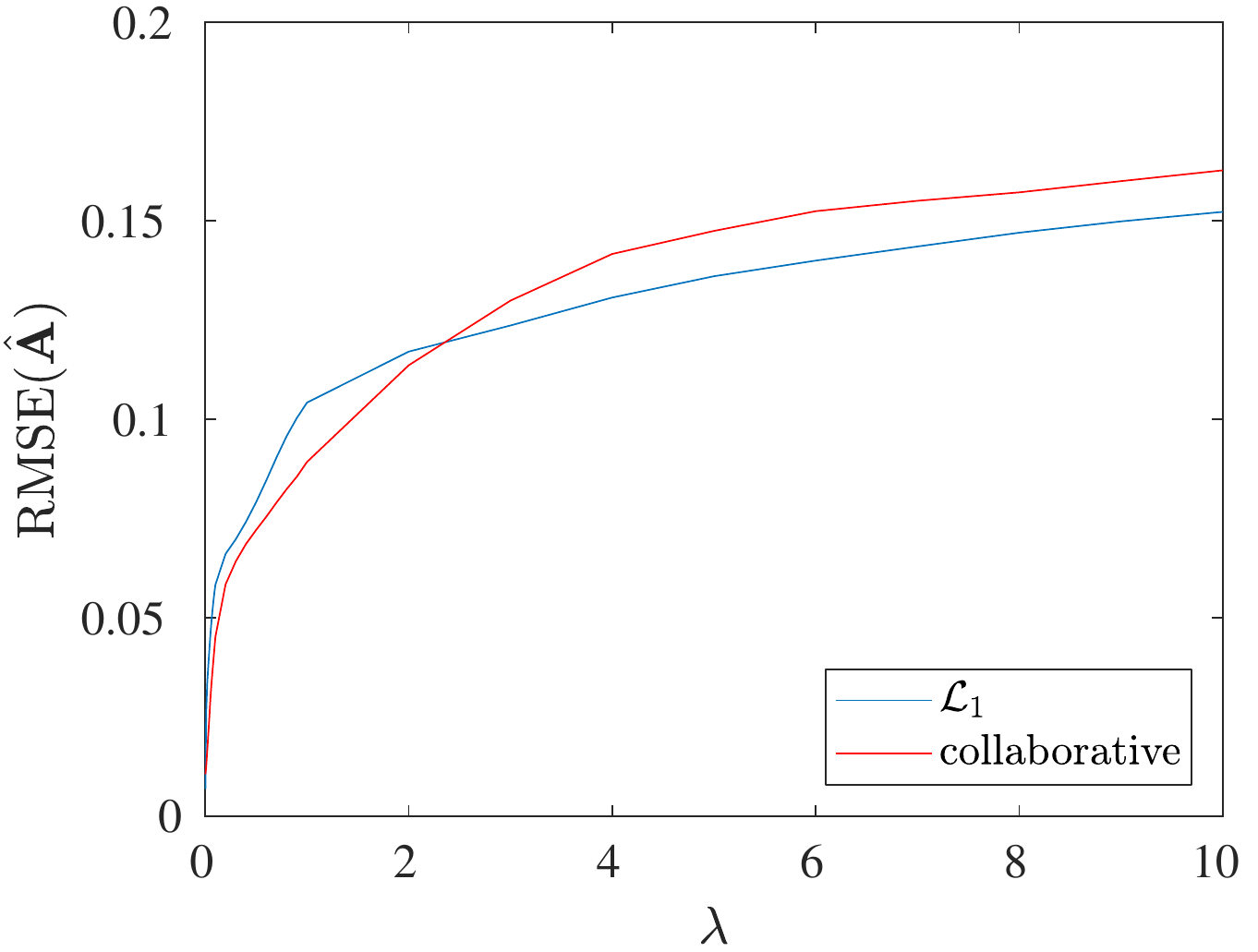}
\caption{RMSE on the abundances on the synthetic data against the regularization parameter $\lambda$ for $\mathcal{L}_{1}$ sparsity (blue), and collaborative sparsity (red).}
\label{sparse_grid_search}
\end{center}
\end{figure}}
%\cc{\begin{figure}
%\begin{minipage}{0.45\linewidth}
%\begin{center}
%\includegraphics[scale=0.3]{grid_search_l1_noasc.pdf}\\
%\centerline{(a)}\medskip
%\end{center}
%\end{minipage}
%\begin{minipage}{0.45\linewidth}
%\begin{center}
%\includegraphics[scale=0.3]{grid_search_collaborative.pdf}\\
%\centerline{(b)}\medskip
%\end{center}
%\end{minipage}
%\caption{RMSE on the abundances on the synthetic data as a function of the regularization parameter $\lambda$ for (a) $\mathcal{L}_{1}$ sparsity, and (b) collaborative sparsity.}
%\label{sparse_grid_search}
%\end{figure}}
From Table~\ref{table_synth}, we see that the elitist algorithm obtains the worst performance in terms of abundance estimation, followed by the FCLSU algorithm, while the group penalty largely outperforms those two when it comes to global abundances. The fractional penalty is able to refine these results slightly and obtains the best performance. If we look at what happens within groups, then the fractional penalty is outperformed by the others. We will see why it is so when we look at the abundance maps. The group and fractional penalties obtain the best endmember RMSE and SAM values, meaning that they recover the closest endmembers to the ground truth. In this configuration, the inter-group sparsity property of the abundances for those two penalties is crucial. The reason the fractional penalty further improves the estimation is that it does not encourage a too harsh within group sparsity, while enforcing inter group sparsity. On the other hand, this sparsity comes at the expense of a precise identification of which endmember candidates were actually used in each pixel. In terms of running time, except for FCLSU which is the fastest, the other approaches are notably slower. The fractional penalty is without surprise the slowest because the constraints in the optimization problem are more complex than for the other two penalties.
\cc{In Fig.~\ref{abs_fractions}, the optimal (over $\lambda$) $RMSE(\hat{\mathbf{A}})$ and $RMSE(\hat{\mathbf{S}})$ are reported for the different fraction values. We see that there is a downward trend for both metrics when the fraction value gets lower. However, the optimal value of $\frac{a}{b}$  for the fractional penalty, among those we tested, varies depending on the considered metric. For $RMSE(\hat{\mathbf{A}})$, it is equal to $0.03$, and for $RMSE(\hat{\mathbf{S}})$ it is equal to $0.01$. A possible explanation is that small values of the fraction make the sparsity inducing term closer to the group $\mathcal{L}_{1,0}$ norm. This shows the interest of the proposed approximation scheme to optimize this mixed fractional norm for any value of $\frac{a}{b}$, instead of simply settling with the cases where $\frac{a}{b} = \frac{1}{2}$ or $\frac{a}{b} = \frac{2}{3}$.  The sensitivity to this parameter is not very critical, as long as it is small enough. Also, we see that the performance values fluctuate more for very low values of the fraction. We attribute this unstability to the fact that the nonconvexity of the penalty gets more and more aggressive, leading to more unstable optimization and possibly to more local minima. In the rest of the experiments, we set $\frac{a}{b} = \frac{1}{10}$.}\\
\cc{\begin{figure}
\begin{center}
\includegraphics[scale=0.25]{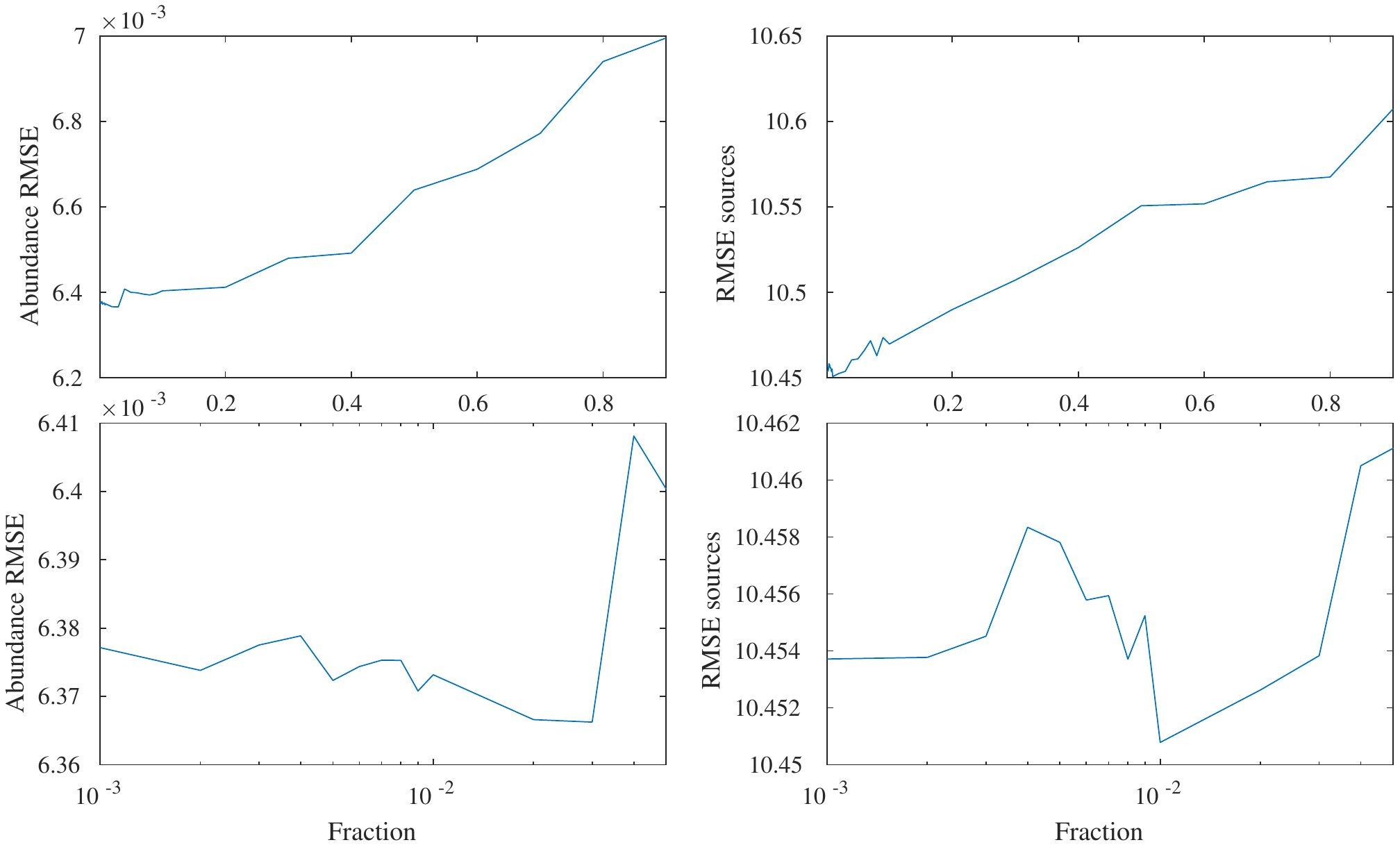}
\caption{Optimal $RMSE(\hat{\mathbf{A}})$ as a function of the fraction used for the fractional penalty. The bottom version is zoomed in for low fractions, with a logarithmic scale on the x-axis.}
\label{abs_fractions}
\end{center}
\end{figure}}
% grid search over fraction too: show that there isn't a big sensitivity
Fig.~\ref{abs_synth_supervised} shows the abundance maps for 8 of the 20 materials to unmix, for all the tested algorithms \cc{except $\mathcal{L}_{1}$ sparsity}. The abundances for the FCLSU algorithm and the group penalty look very similar, with a slight improvement in terms of ``noisiness" of the abundance maps for the latter (e.g. for material 3). \cc{The abundances for collaborative sparsity show a confusion between the groups, because the algorithm tends to null the abundance maps of endmembers which are spectrally similar to others, ignoring the group structure. This confusion happens between materials 3, 5, and 6 in Fig.~\ref{abs_synth_supervised}. These materials all correspond to low brightness endmembers, as can be seen in Fig.~\ref{bundles_synth}.} The abundances for the elitist penalty are slightly worse than the other methods, because although the elitist penalty encourages within group sparsity, it also tends to favor dense mixtures over the groups. The fractional penalty obtains the best visual results, by eliminating most of the noise in the abundances, and getting the support of each material right. However, it seems that it tends to locally make the abundances slightly too close to 1.
\begin{figure}
\begin{center}
\includegraphics[scale=0.43]{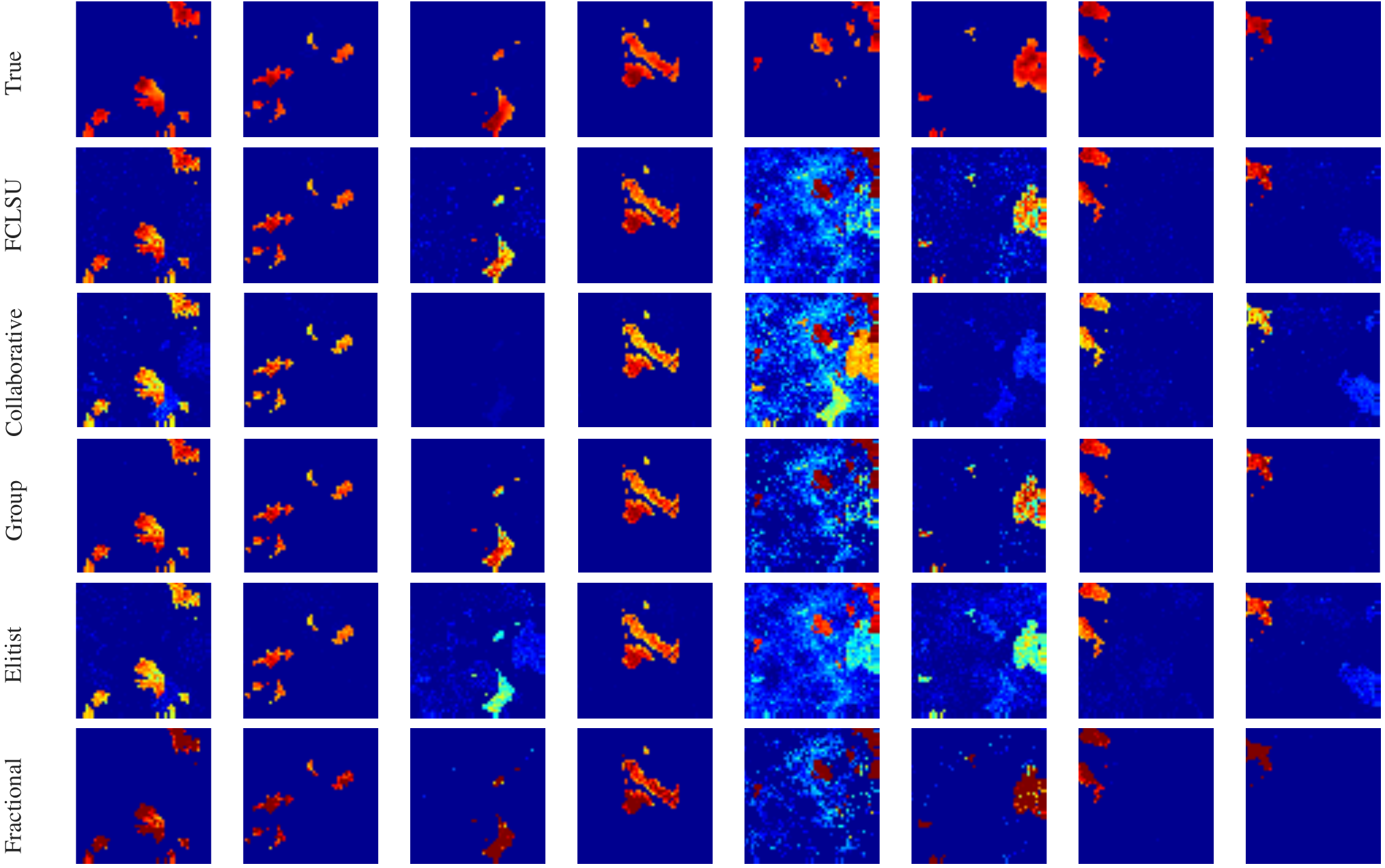}\\
\vspace{0.5cm}
\includegraphics[scale=0.22]{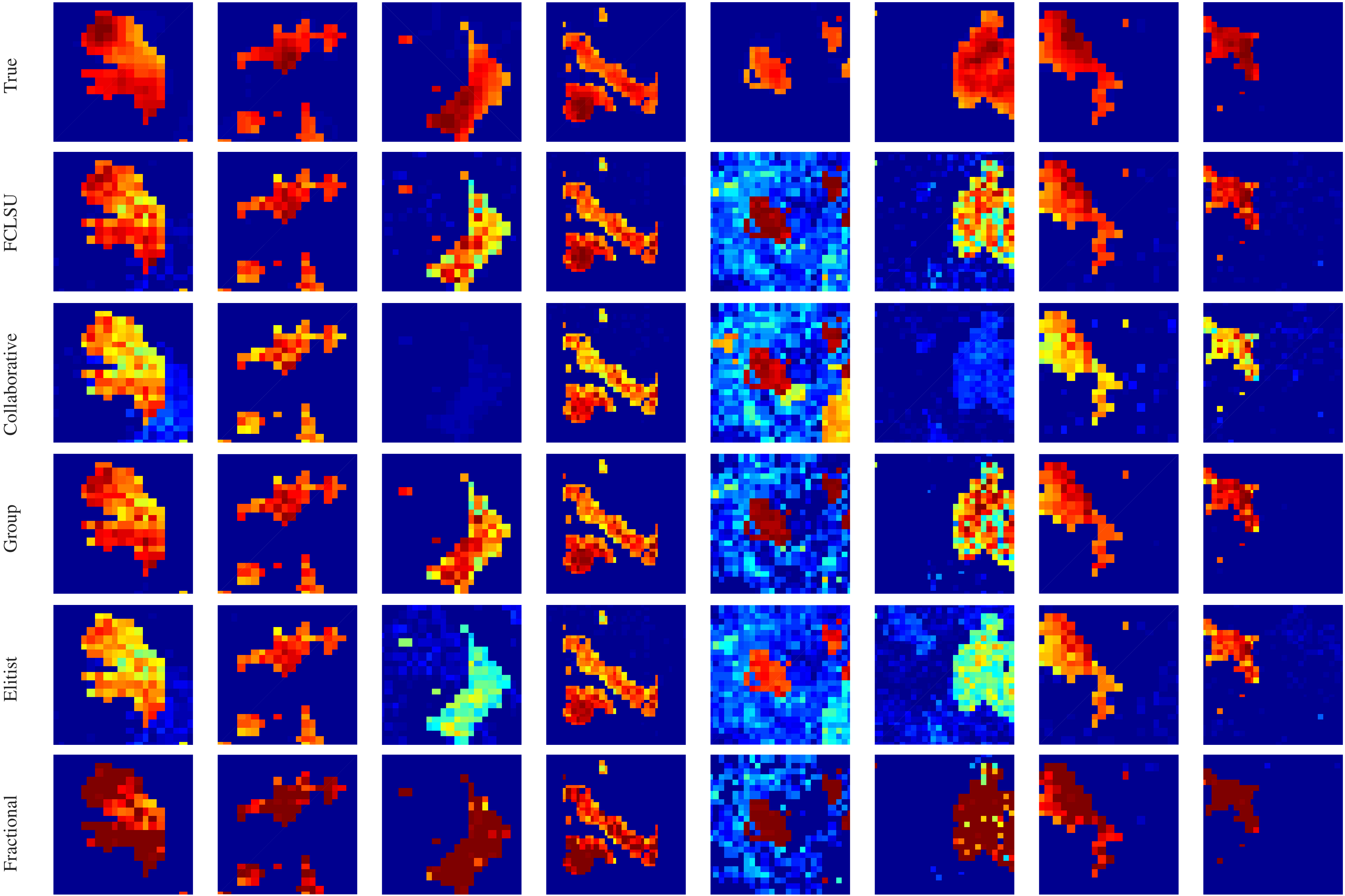}
\includegraphics[scale=0.2]{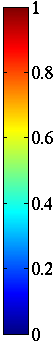}
\caption{Abundances for 8 of the 20 materials, for all the tested algorithms and the ground truth (top rows). The bottom figure is a zoomed version of the abundance maps to better highlight the behavior of the different algorithms.}
\label{abs_synth_supervised}
\end{center}
\end{figure}
%\begin{figure*}
%\begin{center}
%\includegraphics[scale=0.5]{abs_synth_semisup_final.pdf}
%\includegraphics[scale=0.2]{abs_legend.png}
%\caption{Abundances for 8 of the 20 materials, for all the tested algorithms and the ground truth (top row).}
%\label{abs_synth_supervised}
%\end{center}
%\end{figure*}
We also show in Fig.~\ref{abs_matrices_small} a part of the abundance matrices obtained as images (only for some pixels of the image), to show the structure that each penalty induces on the coefficients. We confirm that FCLSU and the group penalty obtain similar performance. Collaborative sparsity deteriorates the abundance estimation because it forces some endmembers to be zero on the whole support of the image. The elitist penalty tends to incorporate spurious endmembers because of the between group density of the coefficients. The fractional penalty obtains a more accurate support of the endmembers than the other two, but at the price of sometimes overestimating the abundances because of a too violent sparse behavior, which is necessary to get rid of the noise but too harsh for a high precision abundance estimation.
\begin{figure*}
\begin{center}
\includegraphics[scale=0.2]{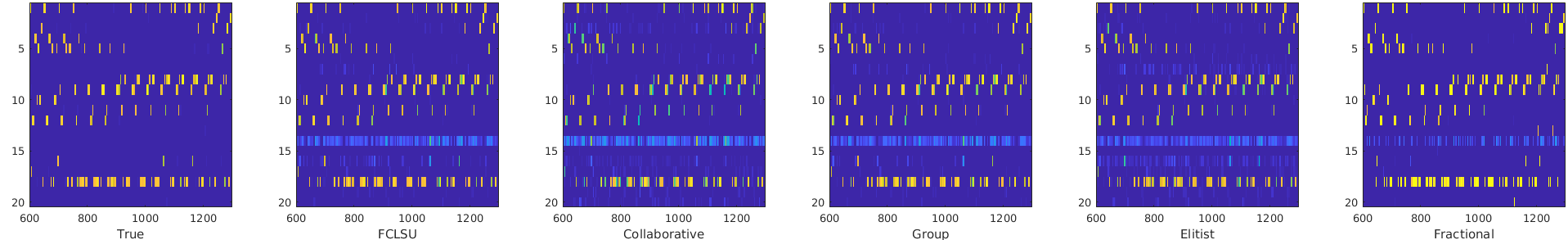}
\caption{Abundance matrices (pixels are on the x-axis, and candidate endmembers are on the y-axis. Candidates 1-5 belong to group 1, candidates 5-10 to group 2, and so on.}
\label{abs_matrices_small}
\end{center}
\end{figure*}
\cc{Finally, we show in Fig.~\ref{bundles_synth} the true endmembers used in the simulation (randomly selected in the USGS spectral library and the ones extracted by the fractional penalty -- the different bundles extracted from all the algorithms are hard to differentiate using  this representation. We can still get some insight on the results from them. First we see that the extracted endmembers interpolate between the candidates present in the bundles, which is in accordance with the geometrical interpretation. Besides, we see that the some endmember signatures with low brightness are very correlated with one another. They correspond to the noisy abundance maps that only the algorithms using the group information, and especially the fractional penalty, are able to unmix.}
%\cc{\begin{figure*}
%\begin{minipage}{0.49\linewidth}
%\begin{center}
%\includegraphics[scale=0.12]{bundles_synth_true.png}\\
%\centerline{(a)}\medskip
%\end{center}
%\end{minipage}
%\begin{minipage}{0.49\linewidth}
%\begin{center}
%\includegraphics[scale=0.12]{bundles_synth_fractional.png}\\
%\centerline{(b)}\medskip
%\end{center}
%\end{minipage}
%\caption{(a) True bundles used for the synthetic data, and (b) bundles extracted by the fractional algorithm. The x-axis is the band number, and the y-axis is the reflectance.}
%\label{bundles_synth}
%\end{figure*}}
\cc{\begin{figure}
\begin{minipage}{0.99\linewidth}
\begin{center}
\includegraphics[scale=0.12]{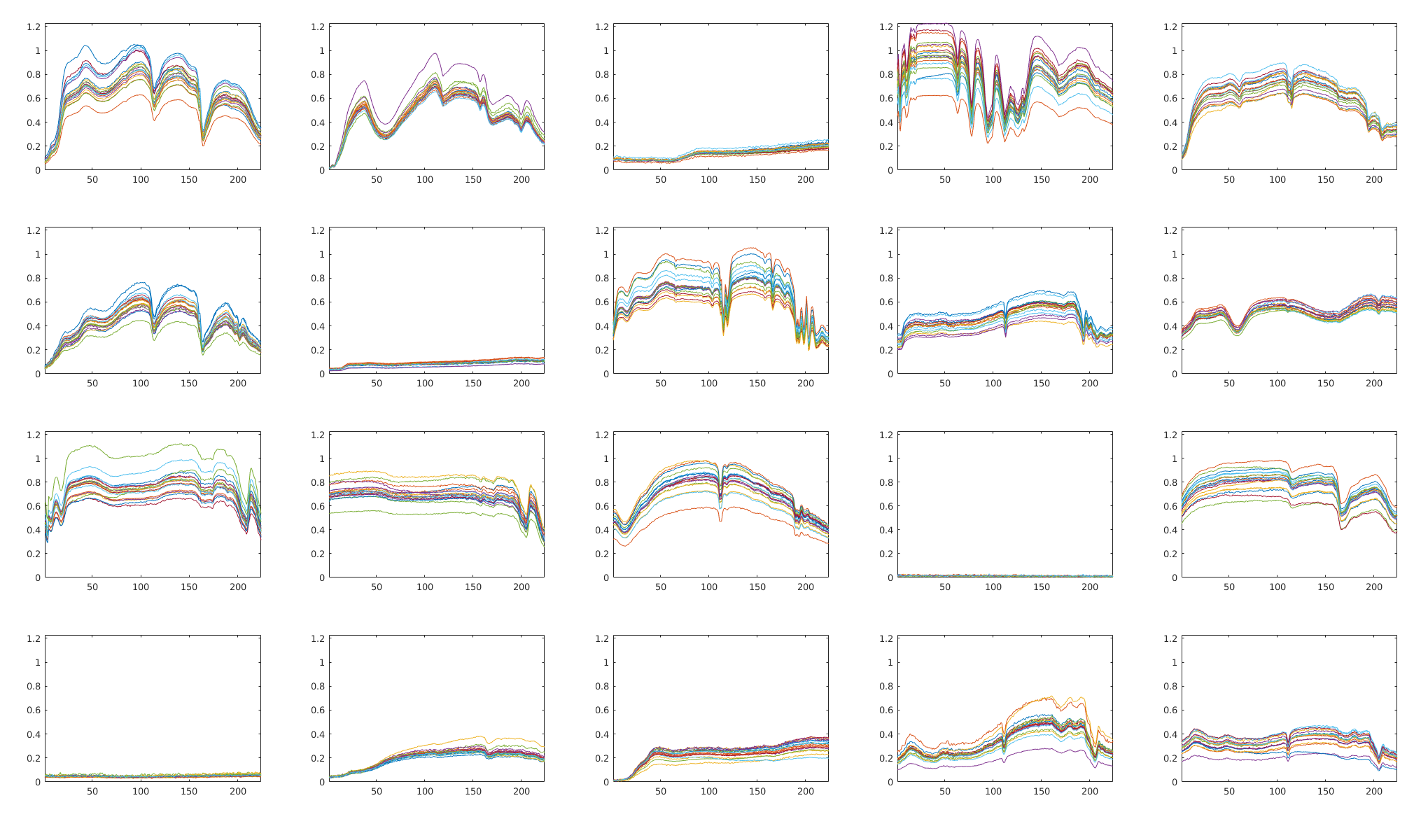}\\
\centerline{(a)}\medskip
\end{center}
\end{minipage}
\begin{minipage}{0.99\linewidth}
\begin{center}
\includegraphics[scale=0.12]{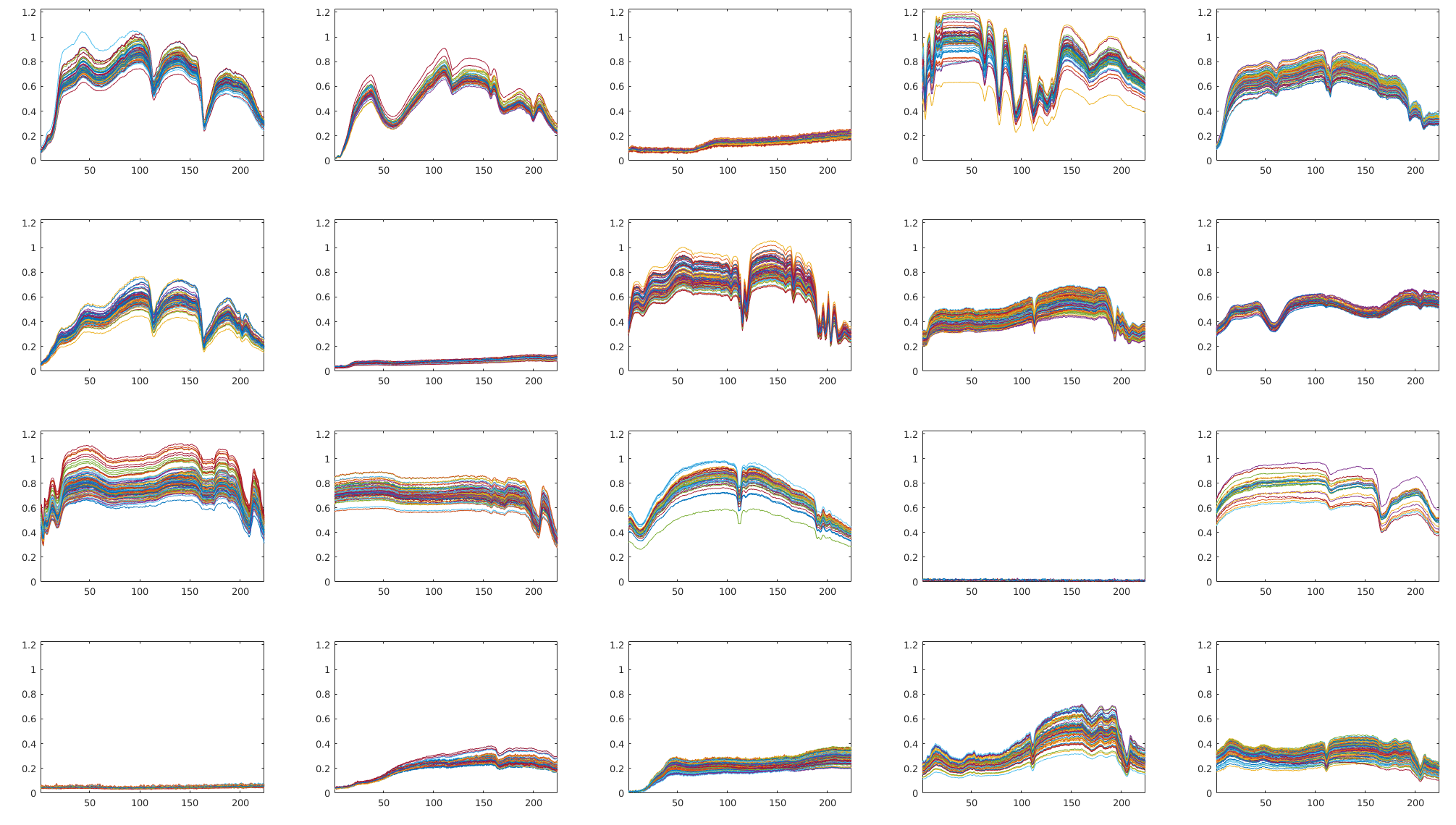}\\
\centerline{(b)}\medskip
\end{center}
\end{minipage}
\caption{(a) True bundles used for the synthetic data, and (b) bundles extracted by the fractional algorithm. The x-axis is the band number, and the y-axis is the reflectance.}
\label{bundles_synth}
\end{figure}}
%Finally, we show in Fig.\ref{SAM_plot} the SAM values for different pixel purities (by only computing the metric in pixels where a material has an abundance higher that a specified theshold) in Fig.~\ref{SAM_plot}, and confirm that variability is better extracted when pixels get purer, and that too low abundances lead without surprise to a bad variability retrieval, as was already noted in~\cite{drumetztip}.
%
%\begin{figure}
%\begin{center}
%\includegraphics[scale=0.4]{SAM_plot.pdf}
%\caption{SAM values as a function of the purity of the pixels considered in the computation (in terms of minimal abundance value for the majoritary material).}
%\label{SAM_plot}
%\end{center}
%\end{figure}

% show abs maps for 4-5 materials (detailed and not), abs matrices, abs tables, and SAM tables. say that SAM gets better when abundance is sufficient (show plot?)
\subsection{Real data}
\subsubsection{Houston Dataset}
\begin{figure}
\begin{center}
\begin{minipage}{0.3\linewidth}
  \centering
\includegraphics[scale=0.18]{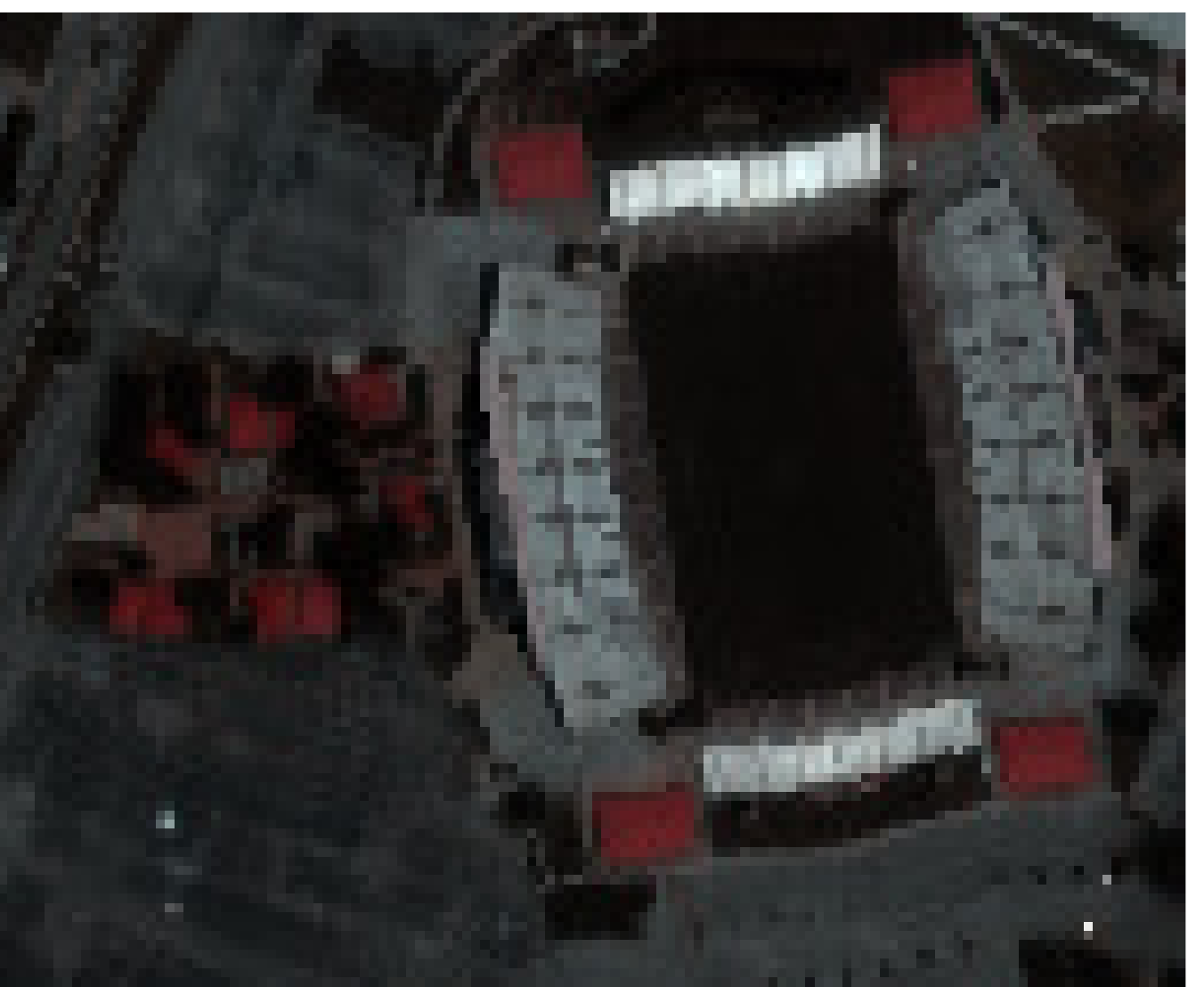}
  \centerline{(a)}\medskip
  \end{minipage}
\begin{minipage}{0.3\linewidth}
  \centering
\includegraphics[scale=0.18]{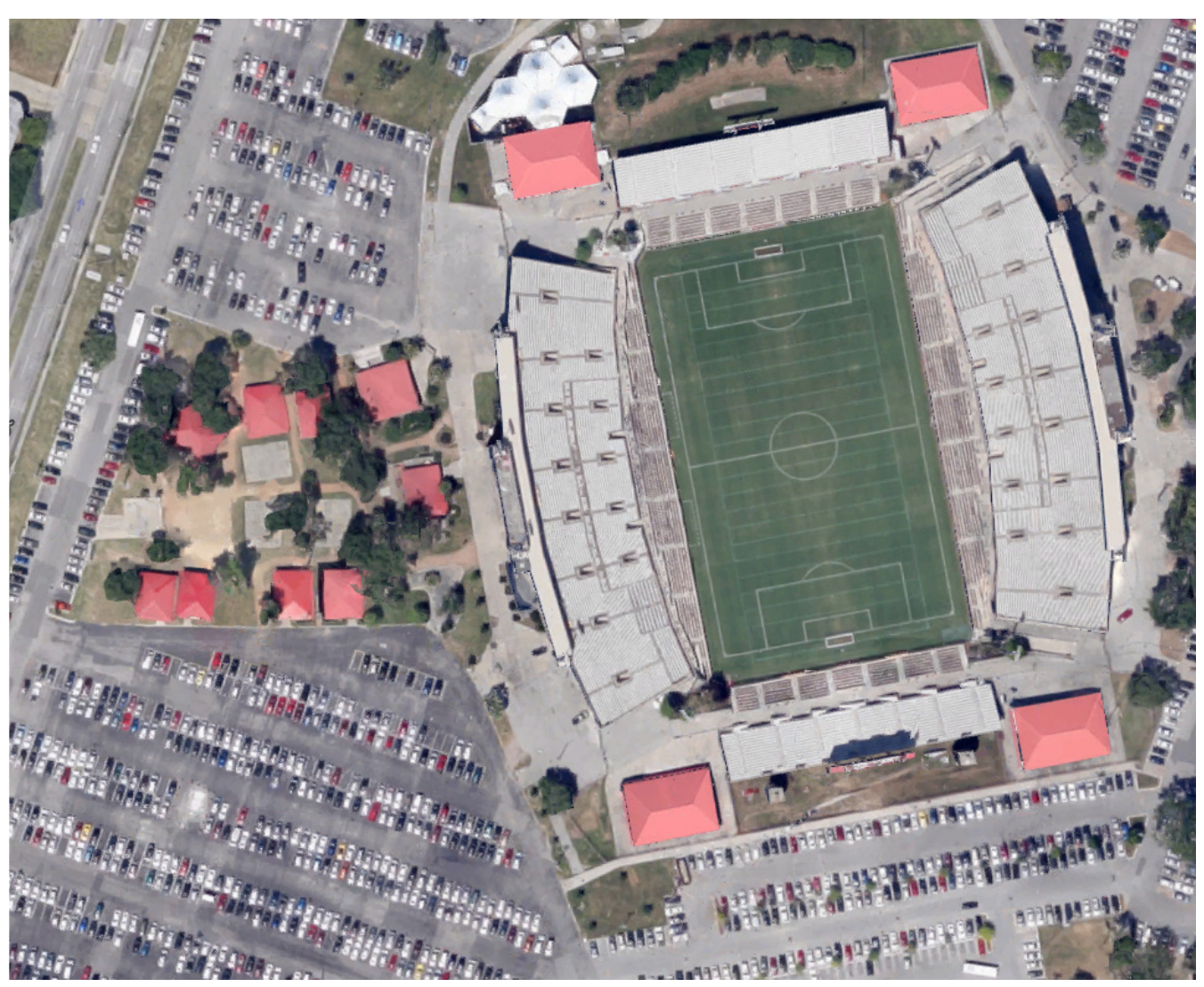}
  \centerline{(b)}\medskip
\end{minipage}
\begin{minipage}{0.3\linewidth}
  \centering
\includegraphics[scale=0.18]{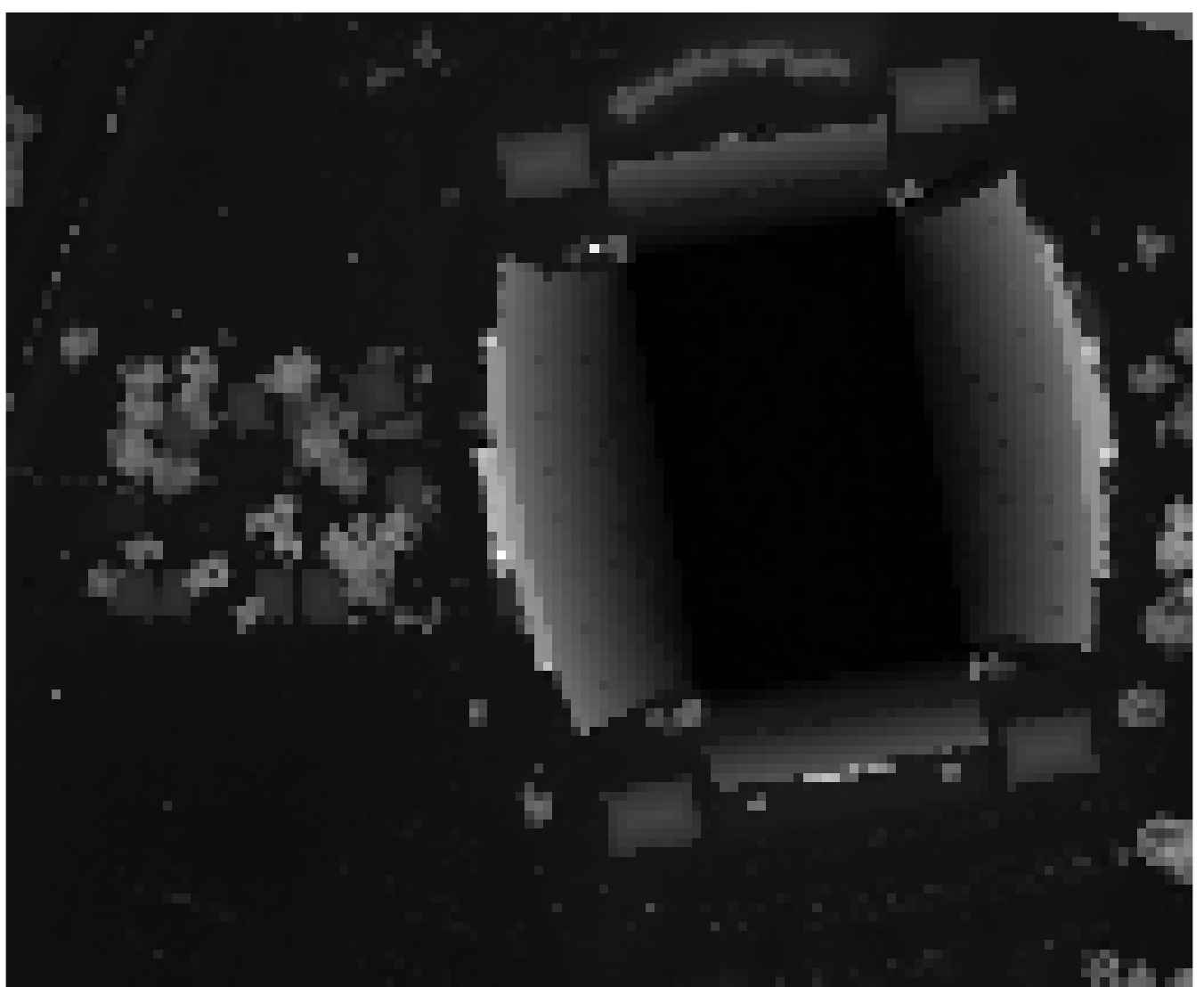}
  \centerline{(c)}\medskip
\end{minipage}
\end{center}
\caption{A RGB representation of the Houston hyperspectral dataset (a). High spatial resolution color image acquired over the same area at a different time (b). Associated LiDAR data (c), where black corresponds to 9.6m and white corresponds to 46.2m.}
\label{Houston}
\end{figure}

\begin{figure}
\begin{center}
\includegraphics[scale=0.25]{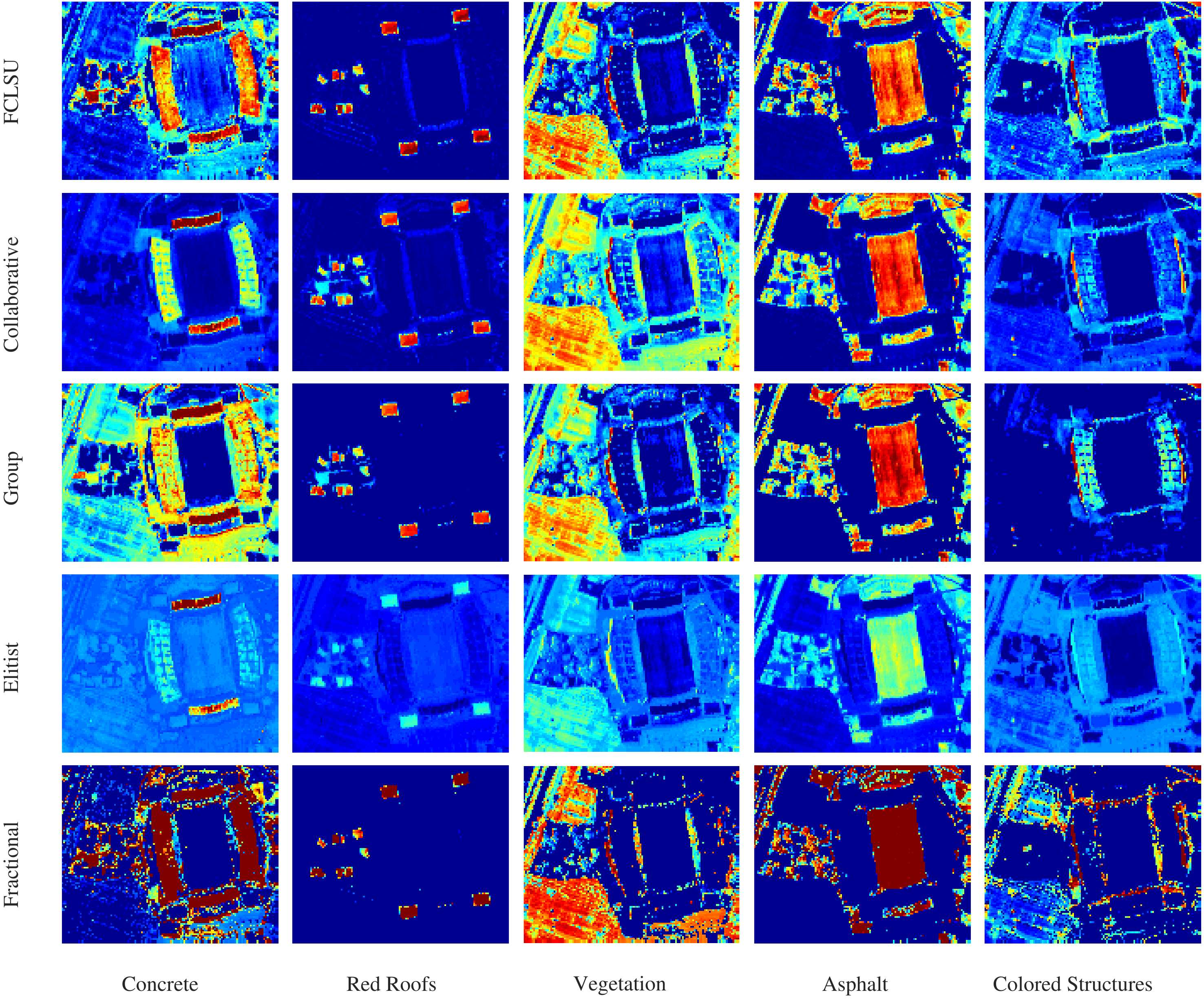}
\includegraphics[scale=0.2]{abs_legend.png}
\caption{Abundances on the Houston dataset.}
\label{abs_real}
\end{center}
\end{figure}

\begin{figure}
\begin{center}
\includegraphics[scale=0.23]{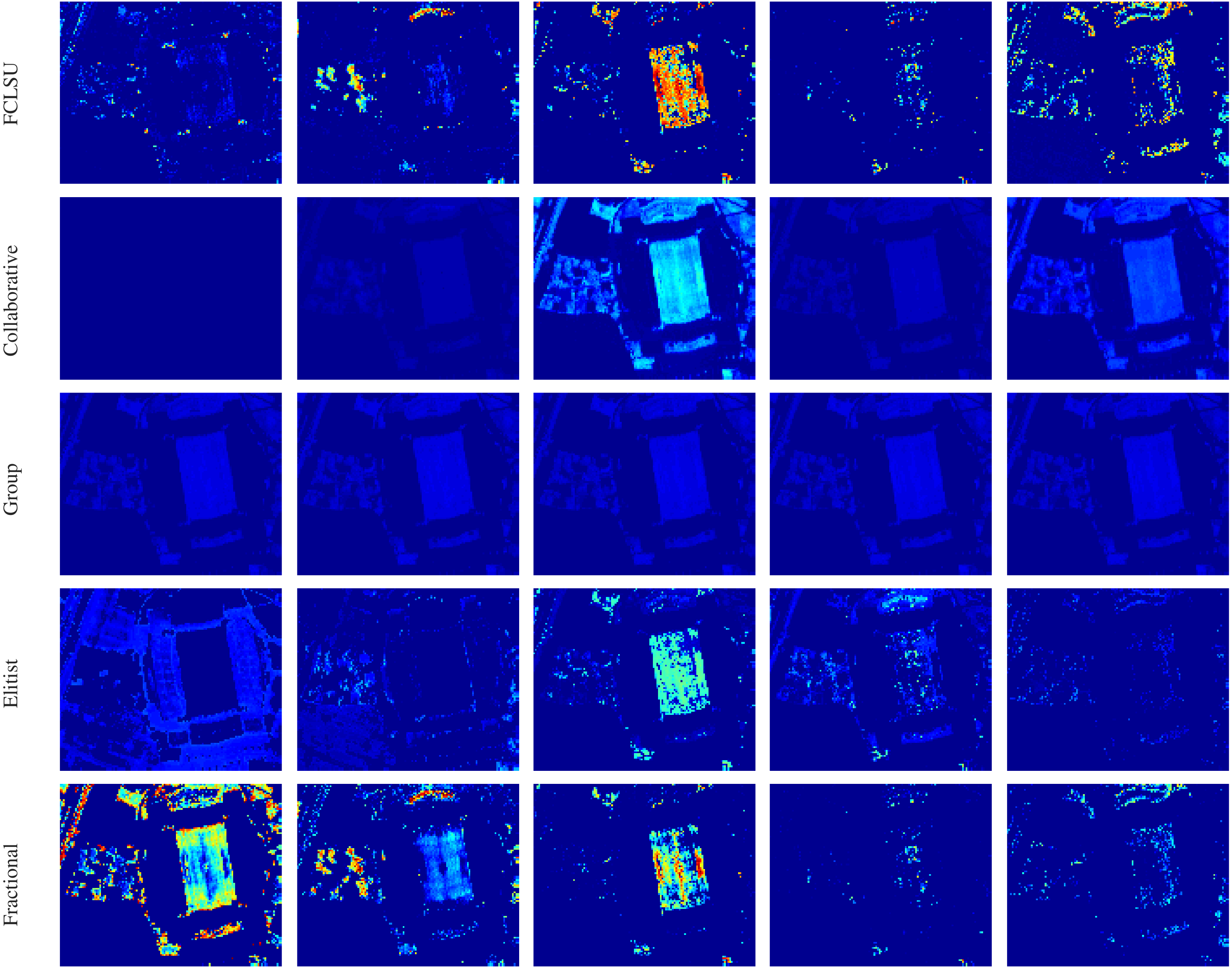}
\includegraphics[scale=0.2]{abs_legend.png}
\caption{Abundances for a few vegetation endmember candidates.}
\label{abs_real_veg}
\end{center}
\end{figure}

\begin{figure}
\begin{center}
\includegraphics[scale=0.24]{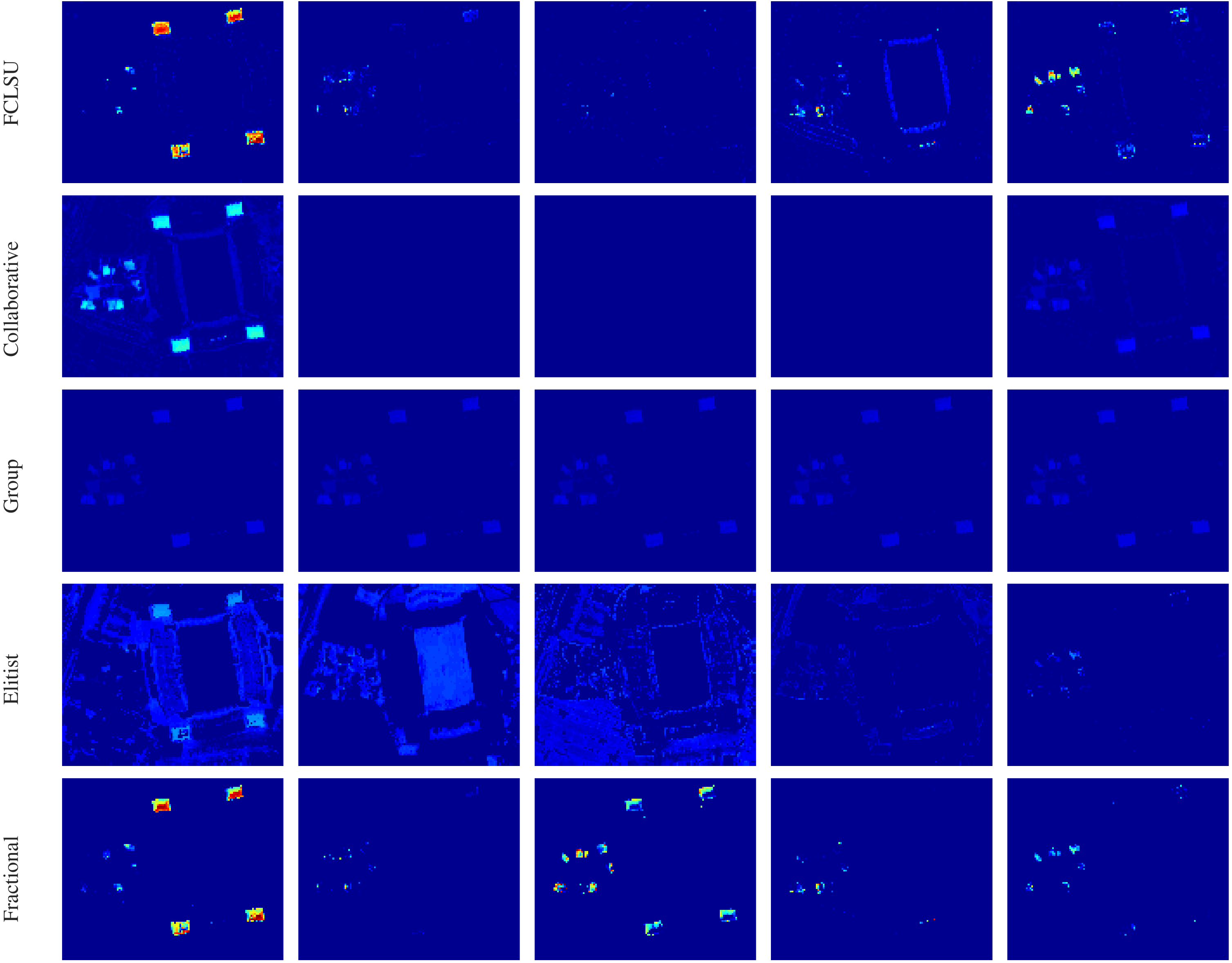}
\includegraphics[scale=0.2]{abs_legend.png}
\caption{Abundances for a few red roofs endmember candidates.}
\label{abs_real_red_roofs}
\end{center}
\end{figure}

\begin{figure}
\begin{center}
\includegraphics[scale=0.2425]{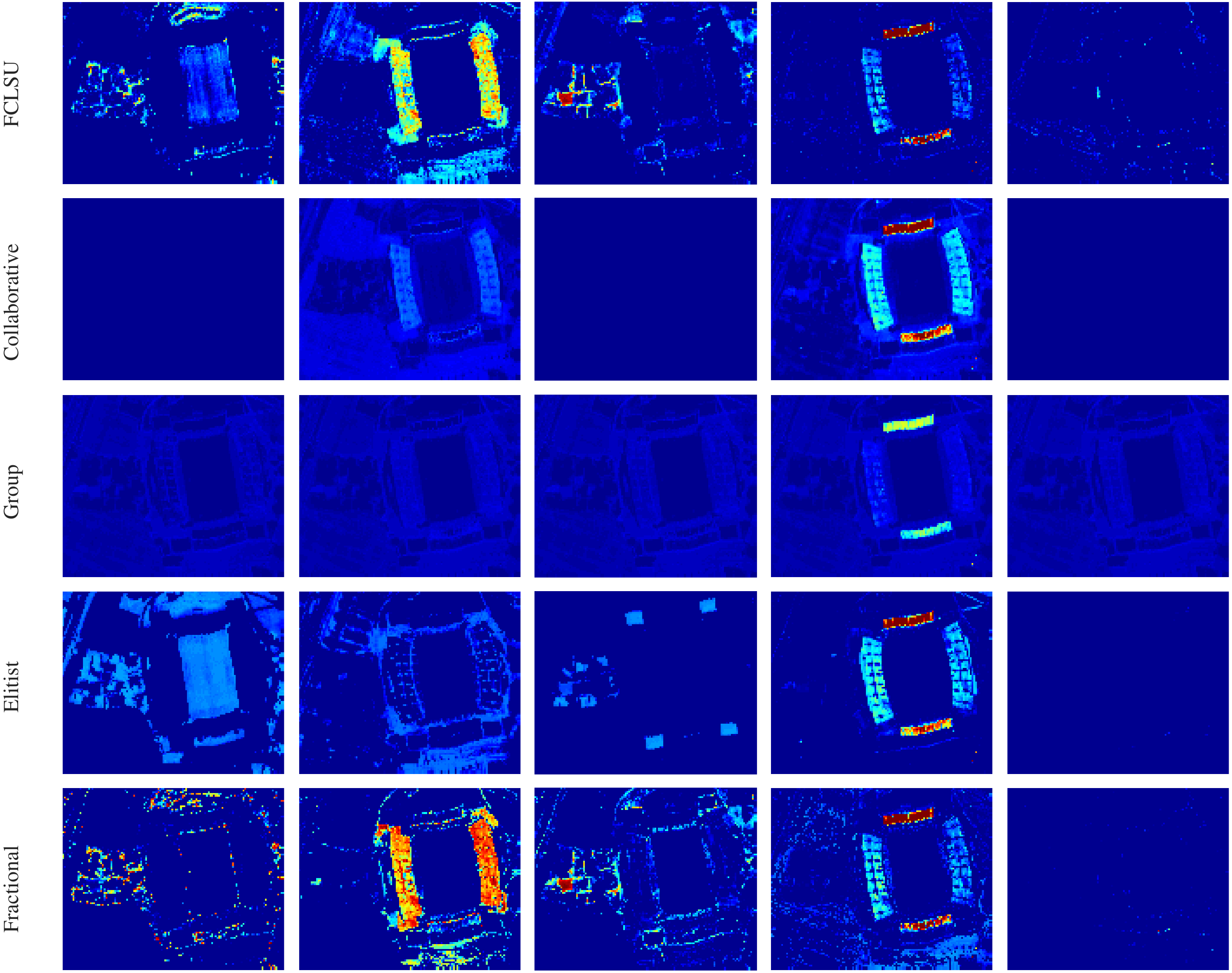}
\includegraphics[scale=0.2]{abs_legend.png}
\caption{Abundances for a few concrete endmember candidates.}
\label{abs_real_concrete}
\end{center}
\end{figure}

The real dataset used in this section is a subset of an image acquired over the University of Houston campus, Texas, USA, in June 2012. It was used in the 2013 IEEE GRSS Data Fusion Contest (DFC)~\cite{Debes2014}. The image comprises 144 spectral bands in the 380 nm to 1050 nm region, and comes with a LiDAR dataset acquired a day before over the same area, with the same spatial resolution (2.5 m). We are interested here in a $152\times 108 \times 144$ subset of this image, acquired over Robertson stadium on the Houston Campus and its surroundings. Fig.~\ref{Houston} shows a RGB representation of the observed scene, as well as a high spatial resolution RGB image of the scene, and a LiDAR-derived digital elevation model. This dataset was used in several works on spectral unmixing and has been shown to exhibit significant variability effects~\cite{drumetztip}. It comprises several structures with locally changing geometries, namely the red metallic roofs which are pyramidal and whose facets correspond to different incidence and emergence angles from the light. The concrete stands also show different orientations with respect to the sun. Vegetation also incorporates variability, since it can be divided into two subclasses: trees and grass (which can in addition can be mixed with soil, which is interpreted as variability in the absence of a soil endmember). \cc{The abundance maps also show that several of these subclasses can be simultaneously active in a single pixel, for instance in the top left corner of the image, where we have pure vegetation pixels with a mixture of trees grass, and soil.}
%\begin{figure}
%\begin{center}
%\begin{minipage}{0.3\linewidth}
%  \centering
%\includegraphics[scale=0.2]{houston_rgb_hyper.pdf}
%  \centerline{(a)}\medskip
%  \end{minipage}
%\begin{minipage}{0.3\linewidth}
%  \centering
%\includegraphics[scale=0.2]{houston_hires.pdf}
%  \centerline{(b)}\medskip
%\end{minipage}
%\begin{minipage}{0.3\linewidth}
%  \centering
%\includegraphics[scale=0.2]{houston_lidar.pdf}
%  \centerline{(c)}\medskip
%\end{minipage}
%\end{center}
Since a library of reference endmembers is unavailable for this image (let alone incorporating variability), the unmixing chain has to be completely blind and we resort to the AEB method to extract bundles. We sample without replacement 10 subsets comprising  10\% of the data pixels each, and extract $P=5$ endmembers on each subset using the VCA. Then we cluster the resulting dictionary into $P=5$ classes using the k-means algorithm (with the cosine similarity measure). Therefore, in this case, we have $Q = 50$. We then compare the performance of the different algorithms in several ways. We use $\frac{a}{b} = \frac{1}{10}$ for the fractional algorithm. The values of the regularization parameters are reported in Table~\ref{table_real}.
\begin{table}
\begin{center}
\scalebox{0.7}{
\begin{tabular}{| c | c | c | c | c | c|}
  \hline 
  \backslashbox{Metric}{Algorithm}   &  FCLSU & Collaborative & Group     & Elitist  & Fractional \\ 
  \hline
 RMSE & \textcolor{red}{0.0029} & 0.010  & \textcolor{blue} {0.005} & 0.016 & {0.008}   \\
  \hline
 SAM (degrees) & \textcolor{red}{0.7240}  & 2.9669 & 2.7891   & 5.1944 & \textcolor{blue}{2.3684}  \\
    \hline
 Running time (s) & 3 & 42 & 8  & 51  & 63  \\
  \hline
 $\lambda$  & $\times$ & 1  & 2  & 0.4   & 0.5  \\ \hline
\end{tabular}}
\end{center}
\caption{RMSE and SAM (computed on the reconstruction) and running times for each tested algorithm on the Houston data. The best values are in red, and the second best are in blue. The regularization parameters are also reported, when applicable.}
\label{table_real}
\end{table}
We first show the abundance maps obtained for each algorithm in Fig.~\ref{abs_real}. From those, we see that all algorithms seem to be able to reasonably explain the material variability present within the image, as opposed to using classical linear unmixing algorithms~\cite{drumetztip}. \cc{Collaborative sparsity obtains visually similar results to FCLSU, except that the stadium is not as well identified.} The group penalty is able to sparsify the abundance maps more than FCLSU \cc{and collaborative sparsity} do, providing clearer abundance maps for the red roofs and the painted structures which are next to the left and right stands of the stadium. However, the detection of the concrete stands themselves is slightly less accurate. The elitist penalty is able to identify the different structures in the image, but fails to consider them as pure because it favors a large number of groups to be simultaneously active in each pixel. The fractional penalty logically provides sparser abundance maps because it penalizes dense mixtures over the groups more aggressively than the group penalty. The structures are considered as purer, which makes sense for an urban scenario such as this one. However, the abundance maps are slightly noisier, with some neighboring pixels in homogeneous regions which do not share the same sparsity properties. This could be mitigated by using spatial regularizations on the abundance maps.\\
Then we show in more detail the abundance maps of 5 of the 10 representatives of the bundles corresponding to vegetation (Fig.~\ref{abs_real_veg}), red roofs (Fig.~\ref{abs_real_red_roofs}), and concrete (Fig.~\ref{abs_real_concrete}). The idea is to find out if the various algorithms are able to provide a meaning to the different representatives within each bundle. \cc{It seems that the collaborative sparsity is able to eliminate a certain number of signatures which are the most redundant, for each material, but tends to distribute abundances relatively equally between the remaining materials, and does not promote pure abundances, which complicate the visual interpretation of the results at the intra class level.} For all materials, the group penalty tends to favor dense mixtures within each group, as expected, making it hard to give an interpretation to each endmember candidate. On the other hand, the elitist penaly obtains much sparser abundance maps within one group, and some endmember candidates can be interpreted (the football field is assigned to a single endmember candidate in the vegetation class, for example). The same goes for FCLSU, which seems to naturally enforce a certain sparsity within each group. Finally, the fractional penalty provides the most conclusive results. For each class, only the predominant signatures are retained and are interpretable, whereas the others are very sparse and some are almost entirely discarded (because they could be spurious, or redundant). For the vegetation class, three subclasses are identified by the algorithm (only two are clearly identifiable for FCLSU): football field, trees, and the area within the field which is mixed with soil. This is interpreted as variability by the model. The other less explicative abundance maps are made more sparse by the algorithm. For the red roof class, the facets of each roof are assigned to two different categories, depending on their orientation w.r.t. the sun, and a third map detects the other red roofs of the image, which are subject to other illumination conditions. The concrete stands are assigned to two different signatures for similar reasons. Note that the abundances of these subclasses are rarely exactly one; this means that the algorithms prefer to identify as a local endmember a convex combination of these two signatures, thus taking into account variability effects which are not sufficiently described by the endmember candidates only.
\cc{To evidentiate the geometrical interpretations of the competing algorithms, we show in Fig.~\ref{scatterplots_houston} the scatterplots corresponding to the different bundles extracted by each algorithm. They reveal the geometric interpretation of the different sparsity inducing penalties. For instance, FCLSU does not promote any type of sparsity, hence the convex hulls of the bundles seem to be relatively uniformly sampled in the results. The collaborative sparsity and the group penalty both lead to local endmembers being situated in the center of the convex hulls. Collaborative sparsity tends to eliminate atoms from the unmixing, but prefers dense mixtures over those which remain. The group penalty prefers dense mixtures within groups to fit the data better. The elitist penalty favors within group sparsity, and we see indeed that more local endmembers are located on the edges of the convex hulls, and less in the center. The fractional penalty exhibits the same behavior, with a more aggressive sparsity, since there are even less endmembers in the center of the convex hulls of the bundles.}
\cc{\begin{figure*}
\begin{center}
\includegraphics[scale=0.22]{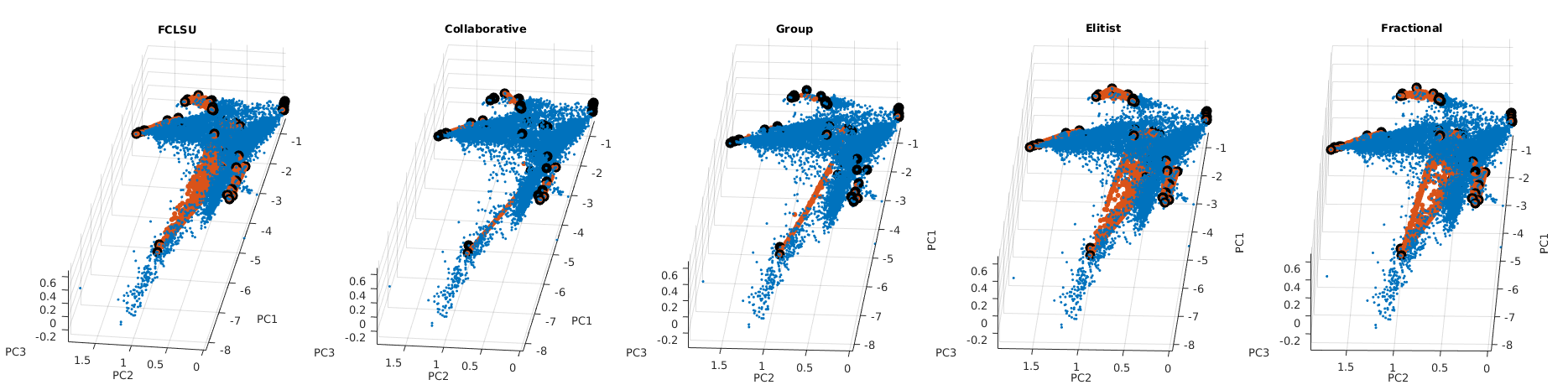}
\caption{Scatterplots (using the first three components of a PCA) of the extracted bundles by each algorithm on the Houston dataset. The data points are shown in blue, the black dots correspond to the extracted bundles, and the red dots corresponds to the local endmembers extracted by the algorithm.}
\label{scatterplots_houston}
\end{center}
\end{figure*}}
Finally, for completeness, we show in Table~\ref{table_real} the overall RMSE and SAM values for each algorithm (computed on the reconstruction of the pixels, in the absence of ground truth), the running times as well as the chosen values for the regularization parameter, when applicable. Without surprise, the RMSE and SAM values decrease when either of the three proposed penalties are used, since we compromise between accurate reconstruction of the data and structured sparsity of the abundance maps. We still note that apart from FCLSU the reconstruction errors are the best when the fractional penalty is used. Since $P$ is rather small here, the running times are more comparable for the different algorithms, except that FCLSU and the group penalty converge after fewer iterations.
%TO DO:\\
%
%DESCRIBE SETUP (BUNDLE CREATION AND SO ON) \\
%
%SHOW GLOBAL ABUNDANCE MAPS FOR ALL ALGS AND COMMENT (more purity with fractional and group to a lesser extent\\
%
%SHOW ABUNDANCE MAPS FOR 5 VARIANTS OF VEGETATION AND ROOFS (fractional gives clearer subclasses and gets rid of unnecessary signatures, while maintaining  very pure abundances).\\
%
%SHOW RMSE AND RUNNING TIMES (lower than for synth because dictionary is much smaller here) \\
%
%WRITE CONCLUSION (new geometric interpretation of unmixing with bundles, proposition of three new penalties to estimate abundances. Group does a better job respecting the structure of the abundances, elitists is penalized by the fact it prefers dense mixtures over the groups, and fractional: mathematically cool, better properties thant using "simpler norms", 1,1/2 or 1,3/5, and allows purer abundances while giving clearer interpretations to subclasses (two level interpretation of the abundance maps).)
\subsubsection{Cuprite dataset}
\begin{figure}
\begin{center}
\includegraphics[scale=0.35]{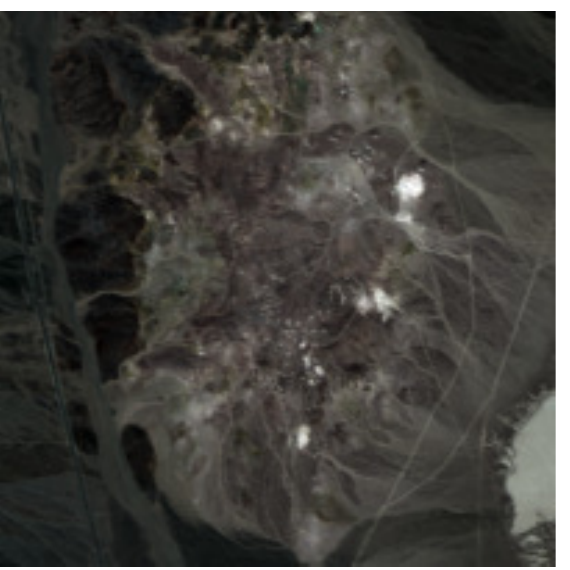}
\caption{\cc{A RGB representation of the subset of the Cuprite dataset used.}}
\label{Cuprite_data}
\end{center}
\end{figure}

\begin{figure}
\begin{center}
\includegraphics[scale=0.28]{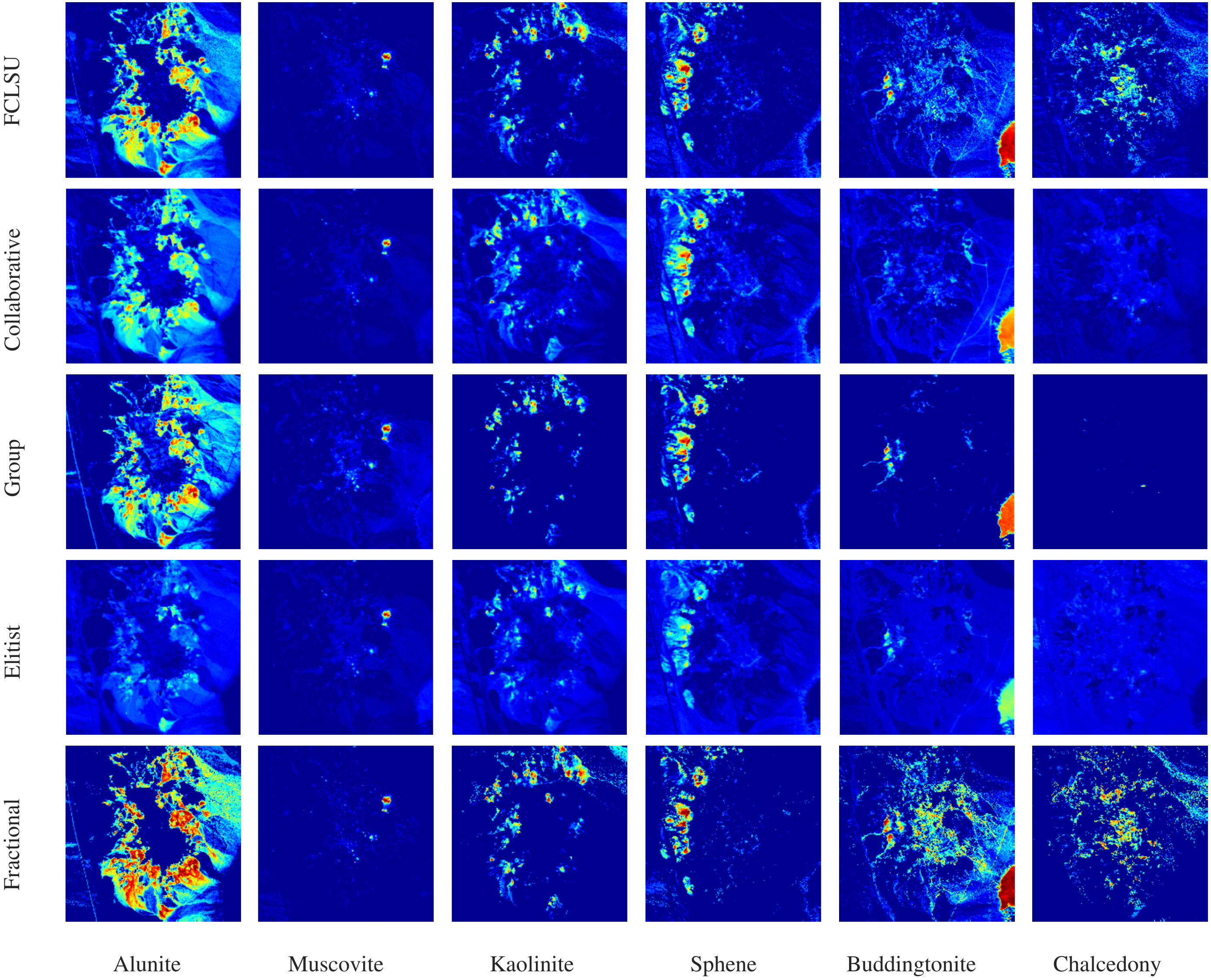}
\includegraphics[scale=0.15]{abs_legend.png}
\caption{\ccc{Abundances on the Cuprite dataset.}}
\label{abs_cuprite}
\end{center}
\end{figure}

\begin{figure}
\begin{center}
\begin{minipage}{0.6\linewidth}
  \centering
\includegraphics[scale=0.22]{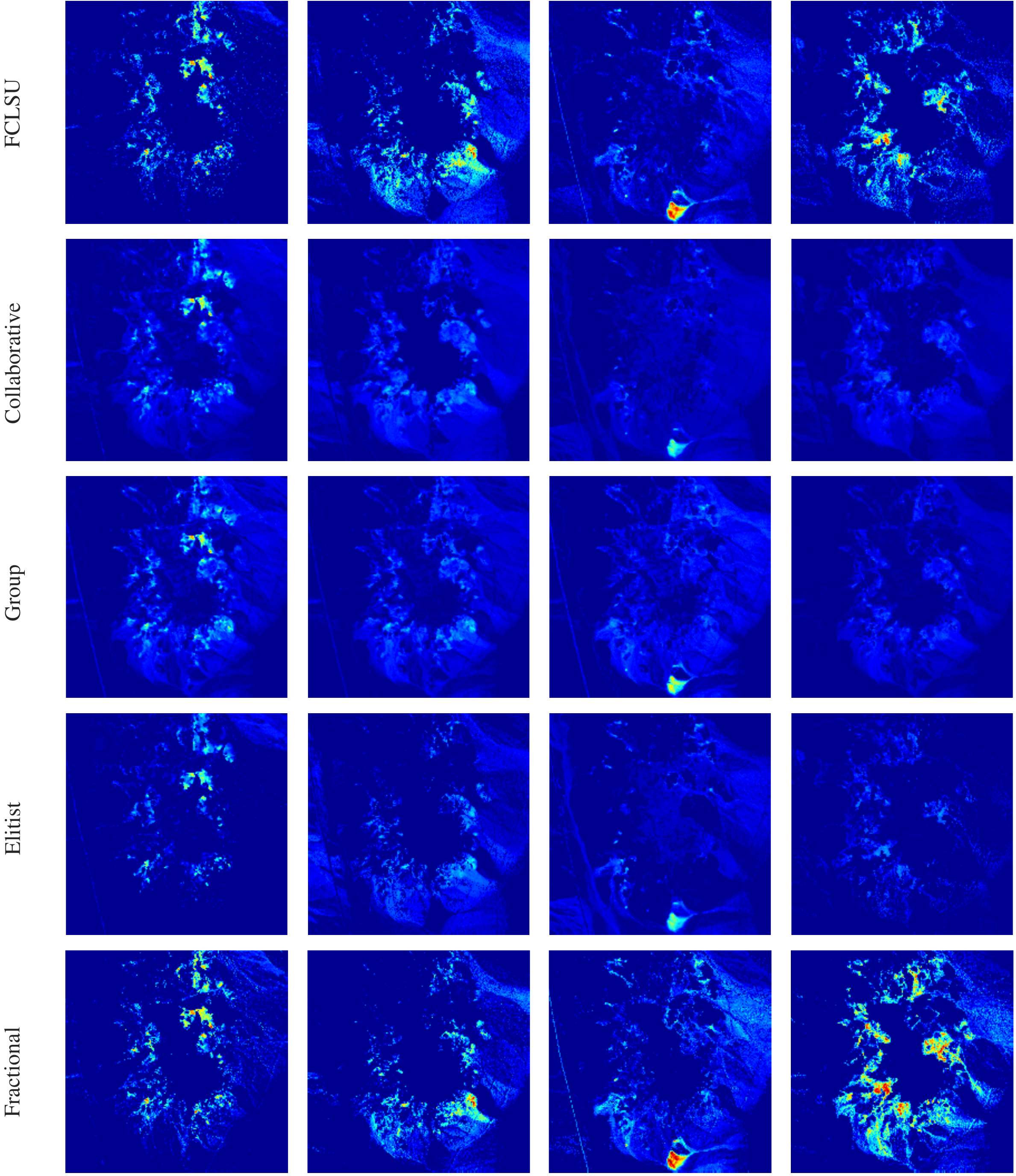}
  \centerline{(a)}\medskip
  \end{minipage}
\begin{minipage}{0.38\linewidth}
  \centering
\includegraphics[scale=0.22]{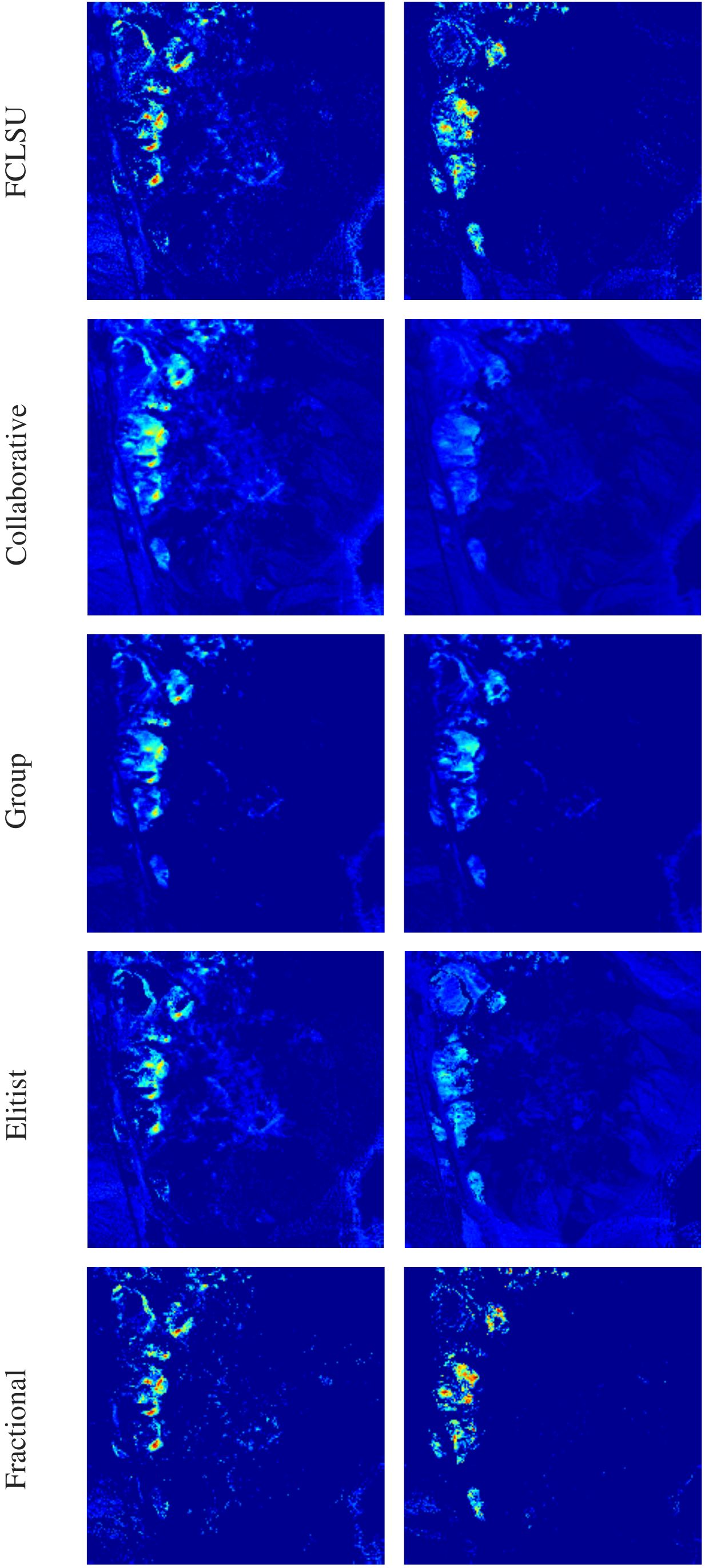}
\includegraphics[scale=0.15]{abs_legend.png}
  \centerline{(b)}\medskip
\end{minipage}
\end{center}
\caption{Abundances for the (a) alunite and (b) sphene endmember candidates.}
\label{bundle_abs_cuprite}
\end{figure}

\cc{The second dataset we consider is a 200 $\times$ 200 $\times$ 186 subset of the Cuprite dataset, which is shown in Fig.~\ref{Cuprite_data}. The image was acquired by NASA's AVIRIS sensor and covers the Cuprite mining district in western Nevada, USA. We extracted 14 bundles according to the intrinsic dimensionality estimated by the Hyperspectral Subspace Identification by Minimum Error (HySIME)~\cite{Bioucas2008} on our subset. We used VCA three times on a third of the dataset, giving a total of $Q = 42$ signatures, clustered into 14 groups as was done for the Houston data. We compare the same algorithms as before and show in Fig.~\ref{abs_cuprite} the estimated abundance maps. The results are shown for only for 6 of the 14 extracted endmembers. The materials have been identified by visual comparison between the estimated abundance maps and endmember signatures with those recovered in~\cite{Nascimento2005}. For two of these materials, i.e. sphene and alunite, we also show the abundances of the different instances of the bundles, in Fig.~\ref{bundle_abs_cuprite}.  The visual results \ccc{(to be compared to a reference land cover map, which can be found e.g. in Fig. 10 of~\cite{li2016robust})} are in accordance with the conclusions drawn from the Houston data, i.e. that the collaborative and group approaches obtain abundances which are sparser and smoother at the global level than FCLSU, because they tend to distribute abundance equally between active signatures in each bundle. The elitist penalty leads to denser mixtures over the groups, and the fractional penalty obtains sparser abundances than other algorithms because it promotes sparsity more aggressively. Also, it obtains within group sparsity, which helps identifying the endmember candidates which are locally predominant, revealing spectral variations for each endmember. \ccc{Some materials, however, are not perfectly separated at the group level, but can be distinguished at the intra group level. Complementary results and their analysis can be found in a supplementary material file provided with this paper.} We also show the RMSE and SAM values associated to the reconstruction of the data, as well as the running times of all algorithms in Table~\ref{table_cuprite}. The numbers follow the trends identified with the Houston data. FCLSU obtains the best quantitative reconstruction results since there is no sparsity involved. The fractional penalty obtains the second best results. The running times are almost equal for the group and elitist penalties, while the fractional algorithm is slightly more computationally intensive.}
%
%
%\begin{figure}
%\begin{center}
%\includegraphics[scale=0.4]{abundances_cuprite.pdf}
%\includegraphics[scale=0.2]{abs_legend.png}
%\caption{\cc{Abundances for all the tested algorithms on the Cuprite dataset.}}
%\label{abs_cuprite}
%\end{center}
%\end{figure}
%\begin{figure}
%\begin{center}
%\includegraphics[scale=0.25]{alunite.pdf}
%\caption{\cc{Abundances for all the tested algorithms on the Houston dataset..}}
%\label{abs_cuprite}
%\end{center}
%\end{figure}
\begin{table}
\begin{center}
\scalebox{0.7}{
\begin{tabular}{| c | c | c | c | c | c|}
  \hline 
  \backslashbox{Metric}{Algorithm}   &  FCLSU & Collaborative & Group     & Elitist  & Fractional \\ 
  \hline
 RMSE & \textcolor{red}{0.0034} & 0.0048  & 0.0075 & 0.0053 & \textcolor{blue} {0.0047}   \\
  \hline
 SAM (degrees) & \textcolor{red}{0.5233}  & 0.7391 & 1.1514  & 0.8288 & \textcolor{blue}{0.7282}  \\
    \hline
 Running time (s) & 54 & 102 & 175  & 174  & 194  \\
  \hline
 $\lambda$  & $\times$ & 1  & 0.05  & 0.1   & 0.125  \\ \hline
\end{tabular}}
\end{center}
\caption{RMSE and SAM (computed on the reconstruction) and running times for each tested algorithm on the Cuprite data. The best values are in red, and the second best are in blue. The regularization parameters are also reported, when applicable.}
\label{table_cuprite}
\end{table}
\section{Conclusion}
\label{conclusion}
In this paper, we have proposed and compared three different sparsity-inducing penalties \cc{to state-of-the-art approaches} specifically aimed at tackling endmember variability in hyperspectral image unmixing, when a dictionary of endmember candidates is available, or when endmember bundles are extracted from the data. We provided a new geometric interpretation of unmixing hyperspectral data within this paradigm. Among the three proposed penalties, the group one enforces a natural inter-class sparsity property of the abundance maps, but is hampered by the fact that it prefers dense mixtures within each group. Conversely, the elitist penalty tries to select the best endmember candidates within each group, but provide inconclusive results because it prefers dense mixtures over the groups. The new fractional penalty we propose leads to the best results: it allows enforcement of both within and inter group sparsity at the same time in a single mixed quasinorm, and is compatible with the sum-to-one constraint on the abundances. The interest of allowing to use any fraction is to provide more flexibility and better results than using the two cases where the shrinkage operators have a closed form ($\frac{a}{b} = \frac{1}{2}$, or $\frac{a}{b}= \frac{2}{3}$). It finally provides purer global abundance maps while giving clearer interpretations to subclasses than the other tested methods. Future work could include the combination of this penalty to spatial regularizations on the abundances, as well as an automatic way to tune the regularization parameter providing a compromise between sparsity and data fit. Finding a proof of convergence of the Alternating Direction Method of Multipliers (ADMM) in the approximation framework we use for the optimization \cc{of the fractional penalty} would also be an interesting perspective.
\bibliographystyle{ieeetr}
\bibliography{biblio}
\vspace{-1cm}
\begin{IEEEbiography}[{\includegraphics[width=1in,height=1.25in,clip,keepaspectratio]{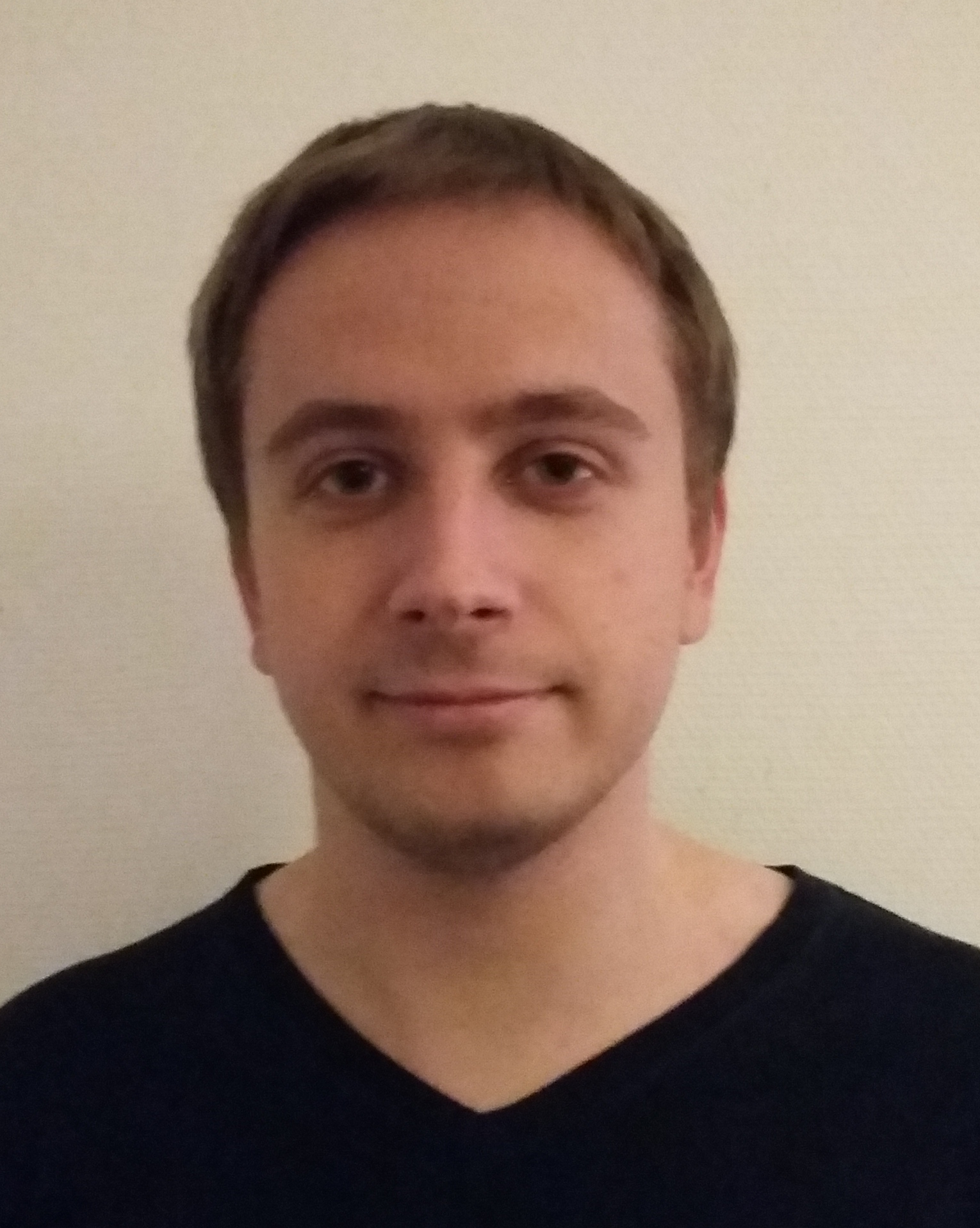}}]
 {Lucas Drumetz} (M’17) received the M.Sc. degree in electrical engineering from the Grenoble Institute of Technology (Grenoble INP), Grenoble, France, in 2013, and the Ph.D degree for the Université de Grenoble Alpes, Grenoble, France, in 2016. He received the 2017 best PhD award from the Université de Grenoble Alpes for this work.  In 2017, he was a visiting assistant professor at the University of California, Los Angeles (UCLA), USA. The same year, he was a visiting researcher at the University of Tokyo, Tokyo, Japan, for three months. Since 2018, he has been an Associate Professor in the department of signal and communications at IMT-Atlantique, Brest, France. His research interests include signal and image processing, inverse problems, optimization techniques and machine learning for remote sensing data.
\end{IEEEbiography}
\vspace{-1cm}
\begin{IEEEbiography}[{\includegraphics[width=1in,height=1.25in,clip,keepaspectratio]{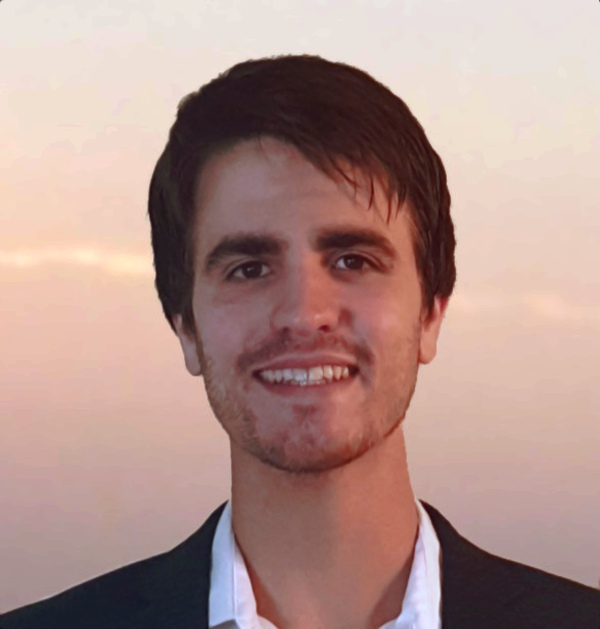}}]
 {Travis R. Meyer} received the B.Sc. degree in physics and the B.A. degree in applied mathematics from the University of California, Los Angeles (UCLA), CA, USA, in 2011. He received his Ph.D. degree in applied mathematics from UCLA in 2017 with focused work on machine learning techniques for text and image analysis. While completing his Ph.D., he helped develop and teach courses on machine learning for his department and mentored many advanced undergraduate students during their first research projects. Additionally, his publications have contributed to areas including atomic force microscopy and hyperspectral imaging. Currently he is the head machine learning developer for SmartKYC where he pursues his interest in variational machine learning techniques by developing cutting-edge knowledge aggregation and analysis systems to help banks understand potential client risks.
\end{IEEEbiography}
\vspace{-1cm}
\begin{IEEEbiography}[{\includegraphics[width=1in,height=1.25in,clip,keepaspectratio]{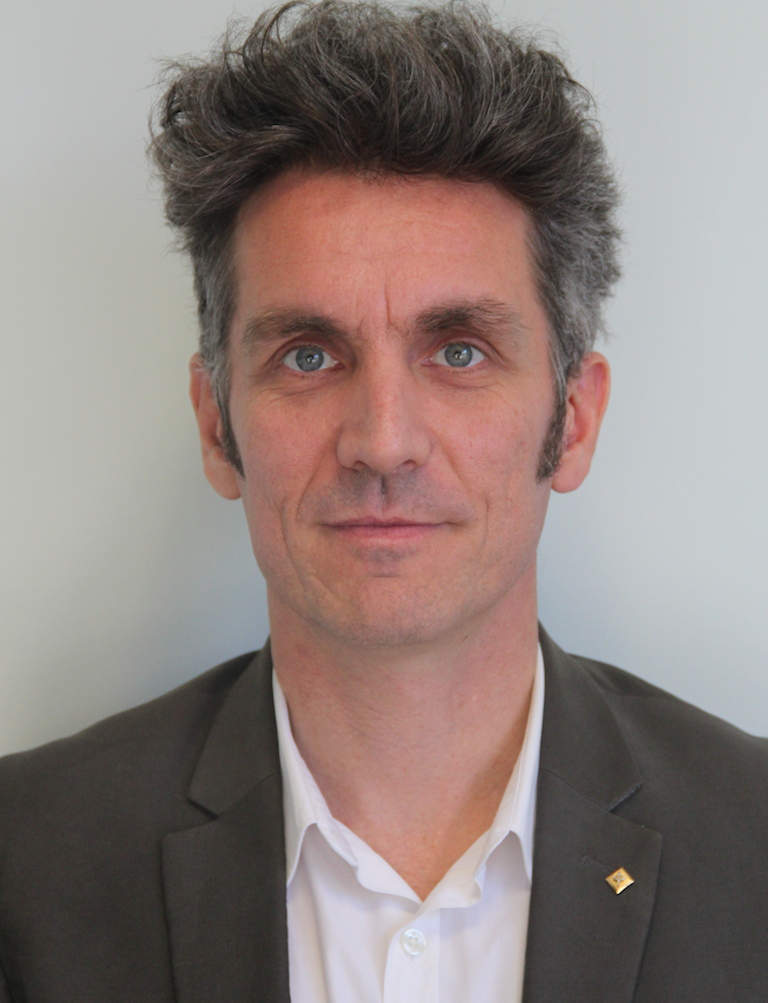}}]
{Jocelyn Chanussot} (M’04–SM’04–F’12) received the M.Sc. degree in electrical engineering from the Grenoble Institute of Technology (Grenoble INP), Grenoble, France, in 1995, and the Ph.D. degree from the Université de Savoie, Annecy, France, in 1998. In 1999, he was with the Geography Imagery Perception Laboratory for the Delegation Generale de l'Armement (DGA - French National Defense Department). Since 1999, he has been with Grenoble INP, where he is currently a Professor of signal and image processing. He is conducting his research at the Grenoble Images Speech Signals and Automatics Laboratory (GIPSA-Lab). His research interests include image analysis, multicomponent image processing, nonlinear filtering, and data fusion in remote sensing. He has been a visiting scholar at Stanford University (USA), KTH (Sweden) and NUS (Singapore). Since 2013, he is an Adjunct Professor of the University of Iceland. In 2015-2017, he was a visiting professor at the University of California, Los Angeles (UCLA).  
Dr. Chanussot is the founding President of IEEE Geoscience and Remote Sensing French chapter (2007-2010) which received the 2010 IEEE GRS-S Chapter Excellence Award. He was the co-recipient of the NORSIG 2006 Best Student Paper Award, the IEEE GRSS 2011 and 2015 Symposium Best Paper Award, the IEEE GRSS 2012 Transactions Prize Paper Award and the IEEE GRSS 2013 Highest Impact Paper Award. He was a member of the IEEE Geoscience and Remote Sensing Society AdCom (2009-2010), in charge of membership development. He was the General Chair of the first IEEE GRSS Workshop on Hyperspectral Image and Signal Processing, Evolution in Remote sensing (WHISPERS). He was the Chair (2009-2011) and  Cochair of the GRS Data Fusion Technical Committee (2005-2008). He was a member of the Machine Learning for Signal Processing Technical Committee of the IEEE Signal Processing Society (2006-2008) and the Program Chair of the IEEE International Workshop on Machine Learning for Signal Processing, (2009). He was an Associate Editor for the IEEE Geoscience and Remote Sensing Letters (2005-2007) and for Pattern Recognition (2006-2008). Since 2007, he is an Associate Editor for the IEEE Transactions on Geoscience and Remote Sensing. He was the Editor-in-Chief of the IEEE Journal of Selected Topics in Applied Earth Observations and Remote Sensing (2011-2015). In 2013, he was a Guest Editor for the Proceedings of the IEEE and in 2014 a Guest Editor for the IEEE Signal Processing Magazine. He is a Fellow of the IEEE and a member of the Institut Universitaire de France (2012-2017).
\end{IEEEbiography} 
\begin{IEEEbiography}[{\includegraphics[width=1in,height=1.25in,clip,keepaspectratio]{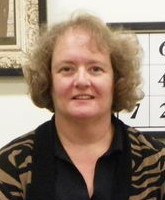}}]
{Andrea L. Bertozzi} is an applied mathematician with expertise in nonlinear partial differential equations and fluid dynamics. She also works in the areas of geometric methods for image processing, crime modeling and analysis, and swarming/cooperative dynamics. Bertozzi completed all her degrees in Mathematics at Princeton. She was an L. E. Dickson Instructor and NSF Postdoctoral Fellow at the University of Chicago from 1991-1995. She was the Maria Geoppert-Mayer Distinguished Scholar at Argonne National Laboratory from 1995-6. She was on the faculty at Duke University from 1995-2004 first as Associate Professor of Mathematics and then as Professor of Mathematics and Physics. She has served as the Director of the Center for Nonlinear and Complex Systems while at Duke. Bertozzi moved to UCLA in 2003 as a Professor of Mathematics. Since 2005 she has served as Director of Applied Mathematics, overseeing the graduate and undergraduate research training programs at UCLA. In 2012 she was appointed the Betsy Wood Knapp Chair for Innovation and Creativity. Bertozzi's honors include the Sloan Research Fellowship in 1995, the Presidential Early Career Award for Scientists and Engineers in 1996, and SIAM's Kovalevsky Prize in 2009. She was elected to the American Academy of Arts and Sciences in 2010 and to the Fellows of the Society of Industrial and Applied Mathematics (SIAM) in 2010. She became a Fellow of the American Mathematical Society in 2013 and a Fellow of the American Physical Society in 2016. She won a SIAM outstanding paper prize in 2014 with Arjuna Flenner, for her work on geometric graph-based algorithms for machine learning. Bertozzi is a Thomson-Reuters/Clarivate Analytics `highly cited' Researcher in mathematics for both 2015 and 2016, one of about 100 worldwide in her field. She was awarded a Simons Math + X Investigator Award in 2017, joint with UCLA's California NanoSystems Institute (CNSI). Bertozzi was appointed Professor of Mechanical and Aerospace Engineering at UCLA in 2018, in addition to her primary position in the Mathematics Department. In May 2018 Bertozzi was elected to the US National Academy of Sciences.
Bertozzi has served on the editorial boards of fourteen journals: SIAM Review, SIAM J. Math. Anal., SIAM's Multiscale Modeling and Simulation, Interfaces and Free Boundaries, Applied Mathematics Research Express (Oxford Press), Applied Mathematics Letters, Mathematical Models and Methods in the Applied Sciences (M3AS), Communications in Mathematical Sciences, Nonlinearity, and Advances in Differential Equations, Journal of Nonlinear Science, Journal of Statistical Physics, Nonlinear Analysis Real World Applications; and the J. of the American Mathematical Society. She served as Chair of the Science Board of the NSF Institute for Computational and Experimental Research in Mathematics at Brown University from 2010-2014 and previously on the board of the Banff International Research Station. She served on the Science Advisory Committee of the Mathematical Sciences Research Institute at Berkeley from 2012-2016. To date she has graduated 35 PhD students and has mentored over 40 postdoctoral scholars.
\end{IEEEbiography} 
\vspace{-1cm}
\begin{IEEEbiography}[{\includegraphics[width=1in,height=1.25in,clip,keepaspectratio]{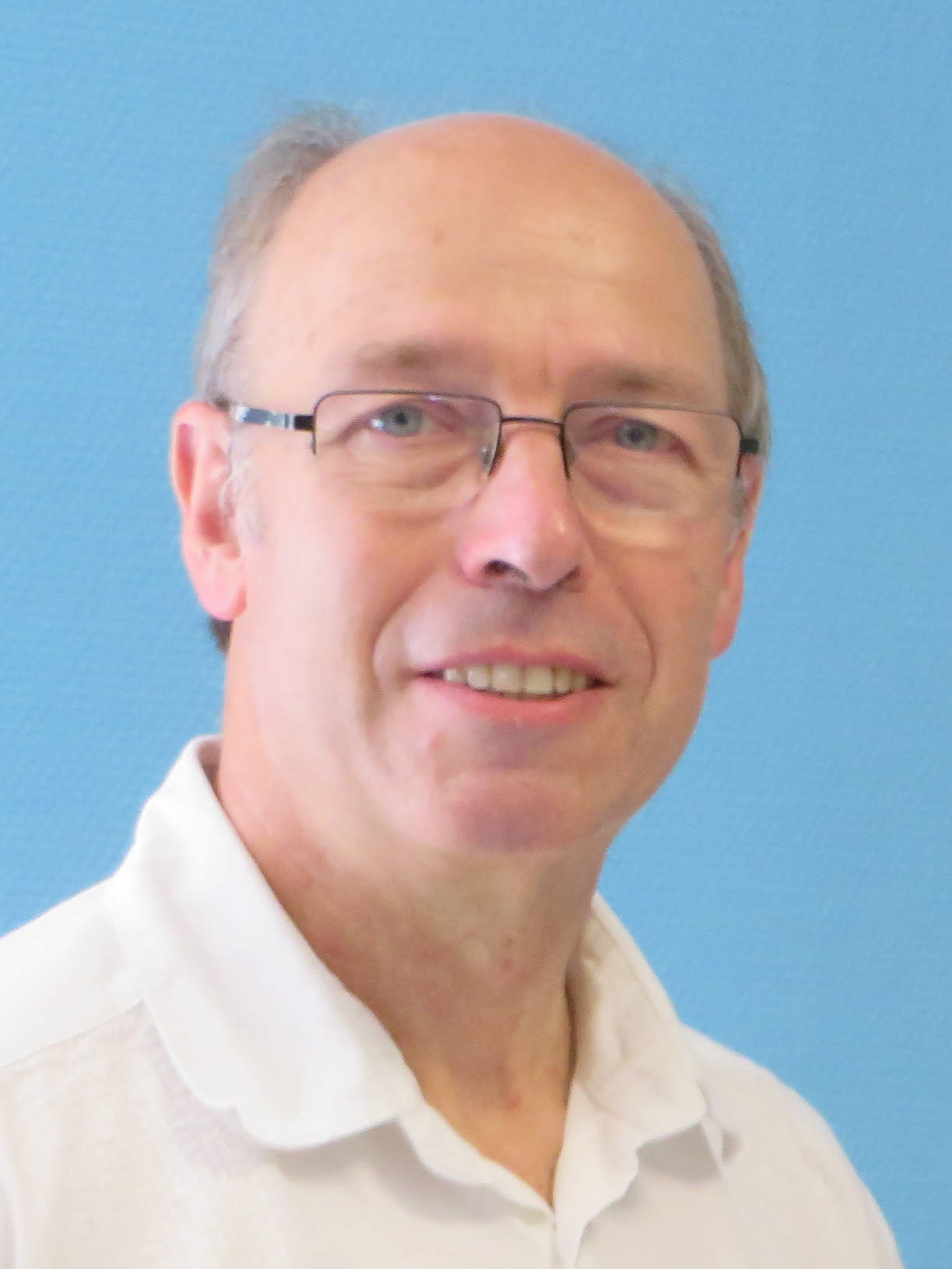}}]
{Christian Jutten} (AM’92-M’03-SM’06-F’08) received Ph.D. and Doctor es Sciences degrees in signal processing from Grenoble Institute of Technology (GIT), France, in 1981 and 1987, respectively. From 1982, he was an Associate Professor at GIT, before being Full Professor at Univ. Grenoble Alpes, in 1989. Since 80’s, his research interests have been machine learning and source separation, including theory (separability, nonlinear mixtures, sparsity, multimodality) and applications (brain and hyperspectral imaging, chemical sensor array, speech). He coauthored 105+ papers in international journals, 4 books, 27 keynote plenary talks and 230+ communications in international conferences. He was a visiting professor at EPFL (Lausanne, Switzerland, 1989), Riken labs (Japan, 1996) and Campinas Univ. (Brazil, 2010). He was director or deputy director of his lab from 1993 to 2010, especially head of the signal processing department (120 people) and deputy director of GIPSA-lab (300 people) from 2007 to 2010. He was a scientific advisor for signal and images processing at the French Ministry of Research (1996–1998) and CNRS (2003–2006 and since 2012). He was organizer or program chair of many international conferences, especially of the 1st Int. Conf. on Blind Signal Separation and Independent Component Analysis in 1999 (ICA’99). He has been a member of a few IEEE Technical Committees. He received many awards, e.g. best paper awards of EURASIP (1992) and IEEE GRSS (2012), Medal Blondel (1997) from the French Electrical Engineering society, and one Grand Prix of the French Académie des Sciences (2016). He was elevated as IEEE fellow (2008), EURASIP fellow (2013) and as a Senior Member of Institut Universitaire de France since 2008. He is the recipient of a 2012 ERC Advanced Grant for the project Challenges in Extraction and Separation of Sources (CHESS). 
\end{IEEEbiography} 

\end{document}